\documentclass[]{jfm}
\usepackage{graphicx}
\usepackage{newtxtext}
\usepackage{newtxmath}
\usepackage{multirow}
\usepackage{natbib}
\usepackage{hyperref}
\hypersetup{
    colorlinks = true,
    urlcolor   = blue,
    citecolor  = black,
}

\newcommand{\RomanNumeralCaps}[1]
\linenumbers
\newcommand{\AW}[1]{{\protect\hypersetup{citecolor=black}\color{black}#1}}
\newcommand{\CSJ}[1]{\textcolor{black}{#1}} 
\newcommand{\EM}[1]{{\protect\hypersetup{citecolor=black}\color{black}#1}} 
\renewcommand{\vec}[1]{\boldsymbol{#1}}

\usepackage{soul}
\newcommand{\Nu}{\mbox{\it Nu}}

\title{
Optimal body force for heat transfer in turbulent vertical heated pipe flow}

\author{Shijun Chu\aff{1},
 Elena Marensi\aff{2}
 \and 
  Ashley P. Willis\aff{1}
  \corresp{\email{a.p.willis@sheffield.ac.uk}}
}

\affiliation{\aff{1} 
Applied Mathematics, School of Mathematical
and Physical Sciences, 
University of Sheffield, Sheffield S3 7RH, UK
\aff{2} 
School of Mechanical, Aerospace and Civil Engineering,
University of Sheffield, Sheffield S1 3JD, UK
}

\begin{document}
\maketitle

\begin{abstract}

\AW{
The vertical heated-pipe is widely used in thermal engineering applications, as buoyancy can help drive a flow, but several flow regimes are possible: shear-driven turbulence, laminarised flow, and convective turbulence.
Steady velocity fields that maximise heat transfer have previously been calculated for heated pipe flow, but were calculated independently of buoyancy forces, and hence independently of the flow regime and time-dependent dynamics of the flow.
In this work, a variational method is applied to find an optimal body force of limited magnitude $A_0$ that maximises heat transfer for the vertical arrangement, 
\EM{ with the velocity field constrained by the full governing equations.}
In our calculations, mostly at $Re=3000$, it is found that streamwise-independent rolls remain optimal, as \EM{in previous steady optimisations},
but that the optimal number of rolls and their radial position is dependent on the flow regime.  Surprisingly, while it is generally assumed that turbulence enhances heat transfer, for the \CSJ{strongly} forced case, time-dependence typically leads to a reduction.
Beyond offering potential improvement through the targeting of the roll configuration for this application, wider implications are that optimisations under the steady flow assumption may overestimate improvements in heat transfer, and that strategies that simply aim to induce turbulence may not necessarily be efficient in enhancing heat transfer either.  Including time-dependence and the \EM{full} governing equations in the optimisation is challenging but offers further enhancement and improved reliability in prediction.
}

\end{abstract}


\section{Introduction}
\label{sec:headings} 
Vertical heated pipe flow is widely used in engineering applications, e.g.\ in geothermal energy capture, nuclear reactor cooling systems and fossil-fuel power plants, to transfer heat from one device to another.   The important difference from iso-thermal pipe flow is that buoyancy, caused by the expansion of the fluid near the heated wall, can partially or even fully drive the flow,  referred to as mixed or natural convection. Mixed convection has been widely researched, due to the interesting and significant effects of buoyancy on the dynamics of the flow and on heat-transfer performance. Buoyancy plays a different role in downward versus upward flow.  In the \AW{former}, buoyancy acts in the opposite direction to the flow and always enhances heat transfer. In an upward flow, buoyancy first deteriorates the heat transfer \citep{ackerman1970pseudoboiling},
then heat transfer recovers only when buoyancy is strong enough. 
Three typical regimes for the heat-transfer characteristics are identified in upward heated flow, namely shear turbulence, laminarised flow, and convective turbulence \citep{parlatan1996buoyancy,yoo2013turbulent,zhang2020review}.  
\AW{(See also figure \ref{fig:SCL}\EM{a}.)}

Extensive research has been conducted to understand the mechanism of heat transfer deterioration in upward heated flow \EM{\citep{hall1969laminarization, steiner1971reverse, carr1973velocity, polyakov1988development, satake2000direct, you2003direct, bae2006effects, jackson2013fluid,he2016laminarisation}.}
\cite{hall1969laminarization} \EM{and \cite{jackson1979influences}} proposed that the reduced shear stress in the buffer layer caused by buoyancy leads to a reduction of turbulence production, suppressing turbulence and even laminarising the flow, consequently deteriorating the heat transfer. 
More recently, \cite{he2016laminarisation}  successfully reproduced the laminarisation phenomenon by modelling the buoyancy with a radially dependent body force added to the isothermal flow. They noticed that the body force causes little difference to the key characteristics of turbulence 
\AW{(in particular, the turbulent viscosity),}
and proposed that laminarisation is caused by the reduction of the ‘apparent Reynolds number’, which is calculated based only on the pressure force of the flow (i.e.\ excluding the contribution from the body force). 
A similar laminarisation phenomenon is also found in isothermal pipe flow, where it has attracted much attention due to its implications for drag reduction \citep{hof2010eliminating,he2016laminarisation,kuhnen2018destabilizing,marensi2019stabilisation}. \cite{kuhnen2018destabilizing} examined the phenomenon of laminarisation from the perspective of the self-sustaining process of shear turbulence \citep{hamilton1995regeneration} and suggested that the decay of turbulence
\AW{can be triggered by modifying the flow such as to reduce}
transient growth. \cite{marensi2019stabilisation} investigated this phenomenon using nonlinear stability analysis \citep{pringle2010using}, and found that nonlinear stability is enhanced in the presence of a body force that flattens the velocity profile. Recently for the vertical heated pipe, \cite{marensi2021suppression} systematically studied the flow regimes and found evidence that heat transfer deterioration and laminarisation are caused by weakened \AW{streamwise vortices}.

Enhanced heat transfer means more effective and efficient energy conversion or cooling, and thus
there have been many interesting investigations aimed at improving the heat transfer in fluid systems.  Strategies can generally be classified as active, passive and compound remedies \citep{webb1983heat,liu2013comprehensive,kumar2015convective,suri2018convective}. Active methods  \citep{ohadi1991heat,wang2020vibration,yuan2023boundary} 
require an external power  input to improve heat transfer. For example,  \cite{ohadi1991heat} studied the effect of corona discharge on forced-convection heat transfer in a tube. \cite{wang2020vibration,yuan2023boundary} proposed a method of vibrating the boundary layer to enhance the heat transfer.  Passive methods include curving or twisting flow geometry, adding extended surfaces and so on. Compound methods \citep{gau1992impingement,naphon2017magnetic,kareem2018mixed} adopt both active and passive techniques. These strategies have significantly improved heat transfer in many systems.

Many techniques for heat transfer enhancement have been developed empirically.  Here we seek a mathematical strategy that is `optimal' with respect to a constrained magnitude of an applied body force.
In principle, maximisation of the heat transfer can be solved by variational methods.  However, the heat-transfer, measured by the Nusselt number $\Nu$,
is a local field variable, depending on the gradient of the temperature evaluated at the wall, which leads to awkward delta functions in a variational approach.  A quantity is required for variational formulations that is spatially global, but which measures the heat transfer at the wall. 
\AW{A suitable quantity is the {\em scalar dissipation} \citep{grossmann2000scaling},
which has a history stemming back to its use 
in the study of mixing of passive scalars
\citep{batchelor1959small}.
In the thermal community, this has recently become known as the entransy dissipation \citep{guo2007entransy,kostic2017entransy}, where it was derived via an analogy between heat conduction and electrical conduction.}
This quantity has been successfully used to optimise heat transfer in various thermal systems, e.g.\ in heat exchangers \citep{guo2012application,guo2010effectiveness} and heat exchanger networks \citep{chen2009optimization}.

For pipe flow, \cite{meng2005field} have  sought a steady velocity field 
that maximises heat transfer.
Although the Navier-Stokes equation was not prescribed as a constraint, it was shown that the velocity field must satisfy a similar equation, subject to a particular force called the synergy force, which produces a velocity field that tends to  align with the temperature gradient \citep{guo2001mechanism}. \cite{jia2014convective} did a similar optimisation but set power consumption as a constraint condition.  They also found that longitudinal swirl flow with multi-vortex structure can enhance heat transfer greatly, and the number of vortexes of the optimal velocity field increases for a larger power consumption.  \cite{wang2015application}  proposed a similar criterion for heat transfer optimisation, exergy destruction minimisation, and a similar optimal velocity field was found.
Such heat transfer optimisations in pipe flow have motivated several heat-transfer enhancement designs, e.g.\ the alternating elliptical axis tube \citep{meng2005field}, discrete double-inclined ribs tubes \citep{li2009turbulent} and many other interesting modifications \citep{liu2013comprehensive,sheikholeslami2015review}. However, the above calculations have assumed a steady laminar flow, while it is common to find that the flow is turbulent.
There have been efforts
to construct variational equations based on the Reynolds Averaged Navier Stokes (RANS) 
turbulence description
 \citep{chen2007field},
 but this approach does not capture the detailed dynamical characteristics of the flow under heating conditions, the self-sustaining mechanisms of the flow and transitions between the flow regimes, 
 that we wish to retain and optimise here.
\cite{motoki2018optimal} also adopted a variational method to find the optimal steady velocity field for plane Couette flow with the largest Nusselt number. 
They found the optimal flow state is composed of streamwise-independent rolls at $Re \sim 10^1$, but there appear smaller-scale hierarchical quasi-streamwise vortex tubes near the walls in addition to the large-scale rolls at $Re \geqslant  10^2$. Although \cite{motoki2018optimal} 
performed calculations 
up to $Re=10^4$, their analysis
assumes a time-independent velocity field.

Optimisations need to be extended to 
\AW{include \EM{both} the 
\EM{momentum} 
equation for the velocity field, in which the buoyancy term may affect the flow regime, and}
time-dependent flows, such as turbulence.
This requires a new framework that includes the dynamical effects of the flow on the mean heat transfer.
The fully nonlinear variational method 
has been used in isothermal pipe flow by \cite{pringle2010using} to find initial 
flow perturbations that grow maximally.
The smallest perturbation which triggers transition is called the `minimal seed'. For a review of wider applications, see  
%
\cite{kerswell2018nonlinear}. In the context of pipe flow, this framework has been successfully employed 
to find the minimal seed under various conditions affecting the flow
\citep{pringle2010using,pringle2012minimal,marensi2019stabilisation}  
and extended to induce transition `the other way', i.e.\ with the aim to construct an optimal `baffle' that destabilises turbulence, thereby causing transition from turbulence to the laminar state \citep{marensi2020designing,ding2020stabilising}. 
The minimal seed for transition in vertical heated pipe flow has been calculated
using the model of \S 2
\citep{chu2024minimal}.
Here, we extend this new nonlinear variational framework to maximise the heat transfer.  While previous optimisations in this geometry have identified optimal stationary velocity fields that maximise the heat transfer, here we seek to optimise a time-independent body force that modifies the time-dependent flow,
which is subject to the full governing equations.
Although hard to accurately reproduce body forces in an engineering application, this is a step towards guiding such an approach.
It should also be noted that the present study mainly focuses on improving heat transfer and understanding the physical mechanism --
changes in the pumping power are not considered, although they are indirectly limited by the amplitude limit we apply to the force.


The plan of the paper is as follows. In \S 2 we present our model of vertical heated pipe flow used for direct numerical simulation (DNS) and the variational equations used in the optimisation.  In \S 3, we first show behaviour for preliminary optimisations, including the features of optimal force, and the effects of variation in the target time. Optimisations are then performed in the laminarisation regime,  shear turbulence regime and convective turbulence regime.  Finally, the paper concludes with a summary in \S 4.

\section{Formulation}

\AW{Consider flow upwards through a vertical pipe that passes through a hot room or chamber.  The fluid will reach the ambient temperature exponentially along the pipe, but will typically do so over sufficiently long distances that the mean temperature can be approximated as linear along subsections of the pipe. The local temperature gradient depends on the rate of heat transfer, and hence on whether the flow is laminar or turbulent.  }

\subsection{The heated pipe flow model}

\begin{figure}
          	\centering
            \includegraphics[angle=0,width=1\textwidth]{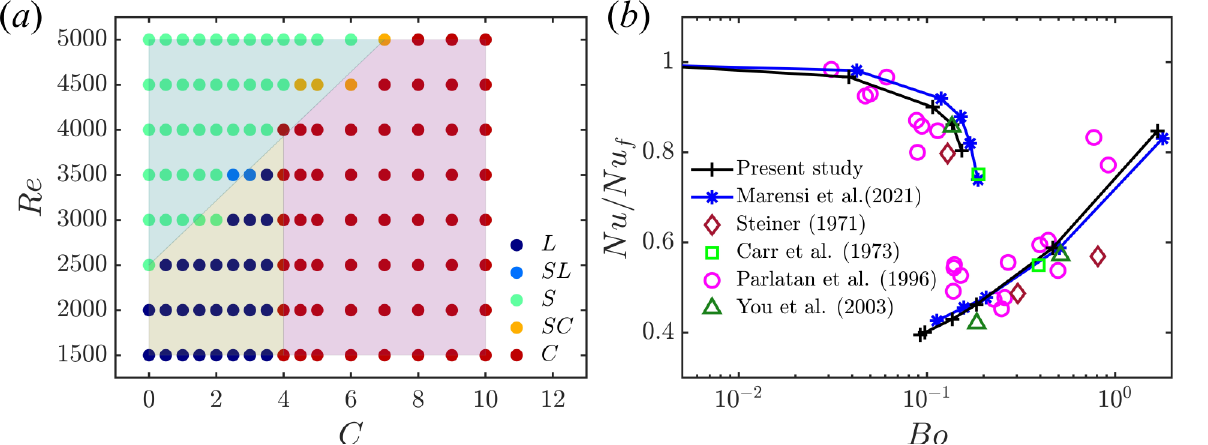}  
          	\caption{ $(a)$ Regions of laminar flow (L), shear  turbulence (S) and convective turbulence (C). SL and SC indicate
           that the flow may be found in either of the two states.
           $\Rey$ and $C$ are the Reynolds number and buoyancy parameters.
           $(b)$ Heat flux measured by $\Nu$, normalised by the value in the isothermal limit ($C\to 0$), as a function of $Bo = 8\times 10^4\,(Nu\,Gr)/(\Rey^{3.425}Pr^{0.8})$. 
           Present data from simulations at $Re = 5300$, $Pr = 0.7$ and various $Gr = 16\,\Rey\,C$. Upper and lower branches correspond to shear turbulence and convective turbulence respectively. 
           \citep[Reproduced from][]{chu2025modelling}
\phantom{\citep{steiner1971reverse,carr1973velocity,parlatan1996buoyancy,you2003direct}}
           }
          	\label{fig:SCL}
     \end{figure}

We begin with the model of
\cite{chu2025modelling}, which assumes a linear mean temperature over the section of pipe considered.  \AW{Axial periodicity of the velocity and temperature fluctuations is also assumed, so that the model should be applied to a section downstream of a bend or inlet effects.} 
A particular feature of the model is that the \EM{mean axial temperature} gradient is allowed to vary in time, reflecting the flow-dependent nature of heat transfer, e.g.\ the flow can transition from the shear-driven turbulent state to the convective state causing a significant drop in the heat flux.
\AW{Good correspondence with previous models and experimental results under statistically steady conditions are shown in figure \ref{fig:SCL}({\it b}), where $Bo$ is an empirically determined buoyancy parameter for which collapse over a large range of parameter sets is observed \citep{jackson2006studies}. }

Using cylindrical coordinates $\vec{x}=(r,\phi,z)$ for a pipe of radius $R$, the total temperature is decomposed as 
   \begin{equation}
   \label{eq:tempexpansion}
    T_{tot}(\vec{x},t)=T_w(z,t) + T(\vec{x},t) - 
    2 T_b
   \end{equation}
where the wall temperature is given by $T_w(z,t)=a_{tot}(t)z+b$, with some constant reference temperature $b$, and  
$T(\vec{x},t)$ carries the temperature fluctuations.
\AW{Heat enters at the hot wall, then exits via a sink term proportional to $a_{tot}(t)$ 
(see (\ref{T-total}) below).}
Let $\langle{\boldsymbol{\cdot}}\rangle$
denote the volume integral and $(1/V)\langle{\boldsymbol{\cdot}}\rangle$ be 
spatial average.
Temperature fluctuations in the model have a fixed positive spatial average $T_b=(1/V)\langle T\rangle$, maintained by adjustments in the temperature gradient $a_{tot}(t)$.  
\AW{As $T_b$ is fixed, the temperature difference between the wall and bulk is constant in the model.}
The constant factor $-2T_b$ has been inserted in (\ref{eq:tempexpansion}) so that the fluctuations $T(\vec{x},t)$ are positive and are largest at the heated wall, 
where $T|_{r=R}=2T_b$.
The heat flux through the wall per unit area is
$q_w=\lambda \overline{\left.({\partial T}/{\partial r})\right|_{r=R}}$,
where $\lambda$ is the thermal conductivity of the fluid and the overline denotes the time average.  
\AW{For laminar flow, 
radial heat flux is purely conductive,
as the flow is purely streamwise. 
The key quantity that measures the observed heat flux through the wall relative to the value for laminar flow} is the Nusselt number
\begin{equation}
    \Nu = \frac{2R\,q_w}{\lambda\,(T|_{r=R}-T_b)} \, .
    \label{eq-Nu}
\end{equation}
\AW{Whilst this definition for $\Nu$ is conventional for pipe flows, it is not precisely $1$ for the laminar case, mostly due to a geometric scale factor. (Also, $T(r)$ and thereby $q_w$ for the laminar case, are not trivially determined, hence the practical dimensional-based approximation.)  However, $\Nu$ will always be presented in the ratio with the value observed in the absence of the body force, denoted $\Nu_{F=0}$, so that such factors drop out.}
We will sometimes use the instantaneous $\Nu(t)$, where the time average is dropped.

Nondimensionalisation using 
the temperature scale $2T_b$, 
the length scale $R$, 
and velocity scale $2U_b$, where $U_b$ is the mean flow speed, and assuming the Boussinesq 
approximation,
leads to the dimensionless governing equations
        \begin{eqnarray}
        	\frac{\partial T }{\partial t}+({\vec{u}_{tot}}\bcdot \vec{\nabla}) T &=&\frac{1}{Re\,Pr}\nabla ^{2}T -{\vec{u}_{tot}}\bcdot{\hat{\vec{z}}}\,a_{tot}(t) \, ,
        	\label{T-total}\\
         \frac{\partial\vec{u}_{tot}}{\partial t}+(\vec{u}_{tot}\bcdot\vec{\nabla}) {\vec{u}_{tot}}&=&-\vec{\nabla} p_{tot} + \frac{1}{Re}\nabla^{2} {\vec{u}_{tot}}+\frac{4}{Re}(1+\beta'(t)+C T )\hat{\vec{z}} \, ,
        	\label{NSE-total}
         \\
       	\vec{\nabla} \bcdot \vec{u}_{tot}&=&0 \, ,
        \end{eqnarray} 
where $\vec{u}_{tot}(\vec{x},t)$ is the velocity field.  The dimensionless boundary condition for the temperature is then 
$T=1$ and the no-slip condition is applied to the velocity, $\vec{u}_{tot}=\vec{0}$, at $r=1$. Axial periodicity over a distance $L$ is assumed for the temperature fluctuations and velocity.
The dimensionless fixed bulk temperature and flow rate are respectively
$(1/V)\langle T\rangle=\textstyle{1/2}$
and $(1/V)\langle \vec{u}_{tot}\cdot\hat{\vec{z}}\rangle=\textstyle{1/2}$, and these two conditions determine values for $a_{tot}(t)$ and the 
excess pressure fraction $\beta'(t)$ via the spatial averages of the respective governing equation.
The Reynolds and Prandtl numbers are $Re=2U_bR/\nu$ and $Pr=\nu/\kappa$, where $\nu$ and  $\kappa$ are the kinematic viscosity and thermal diffusivity respectively.  The third dimensionless parameter 
    \begin{equation}
    C=\frac{Gr}{16\,Re} \, ,
    \end{equation}
    measures the buoyancy force relative to the pressure gradient force that drives laminar isothermal flow, $\mathrm{d}p_{tot}/\mathrm{d}z=4/\Rey$, where 
    $Gr={\gamma g(T|_{r=R}-T_b)(2R)^{3}}/{\nu ^{2}}$ 
    is the Grashof number, $\gamma$ is the coefficient of volume expansion, and $g$ is gravitational acceleration.  
\AW{The parameters $Gr$ and the Richardson number $Ri=Gr/\Rey^2$ are often used in systems involving buoyancy, but in this application the pressure gradient force scales inversely with $\Rey$, and the parameter $C$ is more readily interpreted:
When $C=1$, buoyancy forces are expected to of similar magnitude to the pressure gradient force, and fundamental changes in flow are observed for $C=O(1-10)$.  Figure \ref{fig:SCL}({\it a}) shows the regimes of flow for the model.  The switch at $C\approx 4$ is due to a convection-driven linear instability to helical modes.}  

   
For this configuration, the laminar solution does not have a simple analytic form, and must be computed numerically.  We consider perturbations from the laminar state, subscripted with $0$, and decompose variables as
    ${\vec{u}_{tot}}=\vec{u}_0+\vec{u}$, 
    $p_{tot}=p_0+p$, 
    $1+\beta'(t)=1+\beta_{0}+\beta(t)$, 
    $T=\mathit{\Theta}_0+\mathit{\Theta}$, 
    $a_{tot}(t)=a_{0}+a(t)$.  The 
    perturbations then satisfy
        \begin{eqnarray}
        	\label{T-perturbation}
        	\frac{\partial \mathit{\Theta} }{\partial t}+u_0\frac{\partial \mathit{\Theta} }{\partial z}+u_r\frac{d\mathit{\Theta}_{0}} {dr} + (\vec{u}\bcdot\vec{\nabla}) \mathit{\Theta} &=&\frac{1}{RePr}{\nabla} ^{2}\mathit{\Theta} -u_z a_{0}-(u_0+u_z)a(t)\, , \\
        	\label{NSE-perturbation}
        	\frac{\partial \vec{u}}{\partial t}+u_0\frac{\partial \vec{u}}{\partial z}+u_{r}\,\frac{du_0}{dr}\, \hat{\vec{z}}+(\vec{u}\bcdot \vec{\nabla}) \vec{u} &=& -\vec{\nabla}  p + \frac{1}{Re}{\nabla} ^{2}\vec{u}+\frac{4}{Re}(C \mathit{\Theta}+\beta(t))\hat{\vec{z}}\,,
     \\
       	\vec{\nabla} \bcdot \vec{u}&=&0 \, ,
            \end{eqnarray}
where $\vec{u}=(u_r,u_{\phi},u_z)$.   
Further details on the numerical model can be found in \cite{chu2024minimal} and \cite{chu2025modelling}.

    \subsection{Variational optimisation of a body force for heat transfer}

   In the following, we suppose that a body force $\vec{F}(\vec{x})$ is appended to the right-hand sides of the Navier--Stokes equations (\ref{NSE-total}) and (\ref{NSE-perturbation}), 
   then seek to optimise the form of $\vec{F}(\vec{x})$ such that it maximises $\Nu$,
   subject to a constraint 
   on the magnitude of $\vec{F}$.
   \AW{The {\em scalar dissipation} 
   $\frac{1}{2}\lambda(\bnabla T)^2$ 
   will be used as a proxy for the heat transfer \citep{batchelor1959small,grossmann2000scaling}.  
   (The relationship between $\Nu$ and the scalar dissipation is shown in Appendix \ref{appA}.)
   The term `entransy dissipation' is also being used in thermal science \citep{guo2007entransy,kostic2017entransy}.} 
Here, the time-averaged quantity
    \begin{equation} 
         	\label{eq-J1}
         	J=\frac{1}{\mathcal T}\int_{0}^{\mathcal T}\frac{1}{2} 
            \langle(\nabla T)^2\rangle \, 
            \mathrm{d}t.
    \end{equation}
is used as our objective function.
The way in which the scalar dissipation is used depends upon the thermal boundary condition (Appendix \ref{appA}).
As the fixed temperature boundary condition is applied, maximal $J$ corresponds to maximised `dissipation' of $T^2$ and larger heat flux.  (For the fixed heat flux boundary condition, minimal $J$ corresponds to minimal thermal resistance within the body, and hence maximum heat flux.)

\AW{In this work, we seek for the first time an optimal that is subject to the full governing equations for the velocity field.}  A Lagrangian is constructed as follows:
         \begin{equation} 
         	\label{Larange-heat}
         	\begin{split}
         		\mathcal L=\frac{1}{N}\sum_{i=1}^{N}J_i -\lambda_0\left(\left \langle\frac{1}{2}(\vec{F})^{2}\right \rangle-A_{0}\right)-\sum_{i=1}^{N}\int_{0}^{\mathcal T}\langle\vec{v}_i\bcdot(\mathrm{NS}(\vec{u}_i))\rangle\mathrm{d}t\\- \sum_{i=1}^{N}\int_{0}^{\mathcal T}\langle\Pi_i (\vec{\nabla}\bcdot \vec{u}_i)\rangle \mathrm{d}t - \sum_{i=1}^{N}\int_{0}^{\mathcal T}\langle\pi_i (\mathrm{Tem}(\mathit{\Theta}_i))\rangle\mathrm{d}t \\- \sum_{i=1}^{N}\int_{0}^{\mathcal T}\Gamma_i\langle(\vec{u}_i\bcdot \hat{\vec{z}}) \rangle\mathrm{d}t-\sum_{i=1}^{N}\int_{0}^{\mathcal T}Q_i\langle(\mathit{\Theta}_i) \rangle\mathrm{d}t.
         	\end{split} 
         \end{equation}  
    As the initial condition for the velocity field could be turbulent, to improve the robustness of the results we apply the optimisation to $N$ initial velocity fields.
The variables 
$\lambda_0$, $\Pi_i$, $\pi_i(\vec x,t)$, $\Gamma_i(t)$, $Q_i(t)$ and  $\vec{v}_i(\vec{x},t)=(v_{r, i},v_{\phi, i},v_{z, i})$ are Lagrange multipliers. The first term, the ensemble average of the time-averaged scalar dissipation, is the objective function to be maximised.  The second term fixes the amplitude of the body force; 
\AW{the spatial average of $|\vec{F}|$ is given by $\sqrt{2A_0/V}$.}
Next, the velocity perturbation $\vec{u}$ is constrained to satisfy the Navier--Stokes equation $\mathrm{NS}(\vec{u})$ and the continuity equation,
and the temperature perturbation satisfies the temperature equation $\mathrm{Tem}(\mathit{\Theta})$, each over the period 
from $t=0$ to $t=\mathcal T$. 
The last two terms ensure that the velocity 
satisfies the fixed mass flux and that the bulk temperature is fixed.

     Taking variations of $\mathcal L$ with respect to each variable and setting them equal to zero, we obtain the following set of Euler--Lagrange equations.  The adjoint Navier--Stokes, temperature equation and continuity equations are
     \begin{align}
			\frac{\partial \mathcal L} {\partial \vec{u}_i} = &\frac{\partial \vec{v}_i}{\partial t}+u_0\frac{\partial \vec{v}_i}{\partial z}-v_{z,i}{u_0}'\vec{\hat{r}}+ \vec{\nabla}\times(\vec{v}_i\times \vec{u}_i)- \vec{v}_i\times\vec{\nabla} \times \vec{u}_i +\vec{\nabla} \Pi_i + \nonumber \\ & \frac{1}{Re}\vec{\nabla} ^{2}\vec{v}_i-\pi_i\mathit{\Theta}_{0}'\hat{\vec{r}}-\pi_i\vec{\nabla}\mathit{\Theta}_i-\pi _i (a(t)+a_0(t))\hat{\vec{z}}-\Gamma_i \hat{\vec{z}}=0.
            \label{eq:adjnu}
     \end{align}
	\begin{equation}
		\frac{\partial \mathcal L}{\partial \mathit{\Theta}_i}=\frac{\partial \pi_i}{\partial t}+u_0\frac{\partial \pi_i}{\partial z}+\frac{4}{Re} v_{z,i} C+\vec{u}_i\bcdot \vec{\nabla}\pi_i + \frac{1}{Re Pr}\vec{\nabla}^2\pi_i-Q_i-\frac{1}{\mathcal T }\vec{\nabla}^2 T_i=0 \, ,
        \label{eq:adjT}
	\end{equation}
    \begin{equation}
        \vec{\nabla}\cdot \vec{v}_i=0 \, .
    \end{equation}
       The compatibility conditions (terminal conditions for backward integration
       of (\ref{eq:adjnu}) and (\ref{eq:adjT}) 
       ) are given by
	\begin{equation}
		\frac{\delta \mathcal L}{\delta \vec{u}_i(\vec x,\mathcal T)}=-\vec{v}_i(\vec x,\mathcal T)=0,
	\end{equation} 
        \begin{equation}
    	\frac{\delta \mathcal L}{\delta \mathit{\Theta}_i(\vec x,\mathcal T)}=-\pi_i(\vec x,\mathcal T)=0
        \end{equation} 
	and the optimality condition is
	\begin{equation}
		\frac{\delta \mathcal L}{\delta \vec{F}}=-\lambda_0 \vec{F} +\frac{1}{N} \sum_{i=1}^{N}\int_{0}^{\mathcal T}\vec{v}_i\,\mathrm{d}t =0.
	\end{equation} 
    For an arbitrary initial $\vec{F}$ and set of initial conditions $\vec{u}_i$, the force $\vec{F}$ is incrementally updated to produce a maximum in $\mathcal{L}$
    where
    where $ \delta\mathcal L/\delta \vec{F} $ should vanish. 
    An iterative algorithm similar to that in  \cite{pringle2012minimal} is applied. 
The update for $\vec{F}$ at $(j+1)\mathrm{th}$ iteration is
    \begin{equation}
    \vec{F}^{(j+1)}=\vec{F}^{(j)}-\epsilon_0\frac{\delta \mathcal L}{\delta \vec{F}^{(j)}}.
    \end{equation}
 where $\epsilon_0$ is a small value, controlled using a procedure described in \cite{pringle2012minimal}.  $\lambda_0$ is determined by the condition that $\langle[\vec{F}(\vec{x})^{(j+1)}]^2\rangle=2\,A_0$.
 
\subsection{Numerical methods}
        Calculations are carried out using the open-source code Openpipeflow \citep{willis2017openpipeflow}. Variables are discretised in the domain $\left\{ r,\phi,z \right\}=[0,1]\times[0,2\pi]\times[0,2\pi/\alpha]$, where $\alpha=2\pi/L$, using Fourier decomposition in the azimuthal and streamwise direction and finite difference in the  radial direction, e.g. 
        \begin{equation}
          \label{eq:discretisation}
      	\vec{u}(r_s,\phi,z)=\sum_{k<|K|} \sum_{m<|M|} 
      \vec{u}_{skm}\mathrm{e}^{\mathrm{i}(\alpha kz+m\phi)} \, , \qquad
      s = 1,..., S
    \end{equation}
     where the radial points $r_s$ are clustered towards the wall. Temporal discretisation is via a second-order predictor-corrector scheme, with Euler predictor for the nonlinear terms and Crank-Nicolson corrector.  
     To keep the nonlinear optimisations
     manageable, a Reynolds number $Re=3000$ and $Pr=0.7$ are adopted with a domain of length $L=10$ radii.  We use mesh resolution of $S=64, M=48, K=42$, and the size of the time step is $\Delta t=0.01$.  This resolution is sufficient to maintain a drop-off in the amplitude of the coefficients by three to four orders magnitude, which experience has shown to be sufficient for accurate simulation of shear-driven turbulence.  For the $C$ considered here, the convective state is less computationally demanding to simulate.
\section{Results}
\label{sec:headings} 
  In order to assess the parameters of the optimisation method, 
  we first show
  preliminary optimisations in \S 3.1. Then, we optimise the body force to maximise the heat transfer in the three typical flow regimes of vertical heated pipe flow,
  i.e.\ the laminarisation regime (\S 3.2), shear turbulence regime (\S 3.3) and convective turbulence regime (\S 3.4).
  \EM{(Parameter regimes for this model are shown in figure \ref{fig:SCL}.) }
  
  \subsection{Preliminary optimisation} 

   For the laminar case, the state is unique and  we require
   only one initial velocity field, $N=1$.
   We start with the unforced laminarised flow at $\Rey=3000$, $C=3$, 
   and take random fields for the initial force (such as a turbulent velocity field).
        \begin{figure}
         	\centering
            \includegraphics[angle=0,width=1\textwidth]{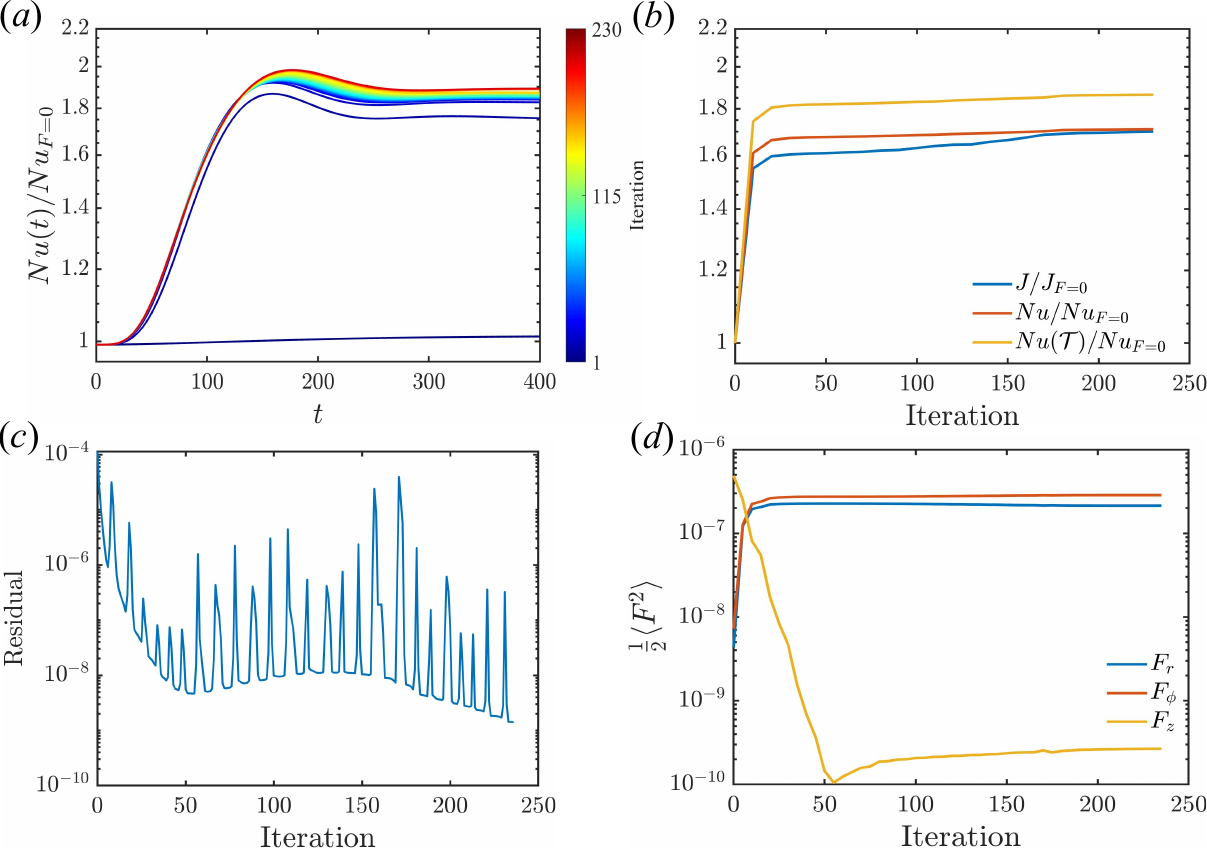}  
         	\caption{ $(a)$ Instantaneous Nusselt number $\Nu(t)$ as the iteration proceeds, where $\Nu_{F=0}$ refers to the value of the unforced case. $(b)$ 
          Dimensionless time-averaged scalar dissipation, (time-averaged) $\Nu$, and $\Nu(\mathcal T)$
versus iteration, normalised by their values at the zeroth iteration. $(c)$ The residual $\langle({\delta \mathcal L}/{\delta \vec F})^2\rangle$ versus iteration. $(d)$ Magnitude of the  components of body force $\frac{1}{2}\langle F_r^2\rangle,
          \frac{1}{2}\langle F_\phi^2\rangle, 
         \frac{1}{2} \langle F_z^2\rangle$ 
          versus iteration. 
     The optimisation is run at $A_0=5\times10^{-7}, \mathcal T=400, C=3, Re=3000$. }     
         	\label{prelical}
        \end{figure}
Results from a preliminary optimisation with $A_0=5 \times 10^{-7}, \mathcal T=400$ 
   are shown in figure \ref{prelical}.  The instantaneous Nusselt number, $\Nu(t)$,
normalised by the mean for the unforced flow, $\Nu_{F=0}$, is shown in figure \ref{prelical}$(a)$ for each iteration.  The final value increases by more than 80\% over the unforced laminar case.  Figure \ref{prelical}$(b)$ shows the objective function $J$ \eqref{eq-J1}, the (time-averaged) Nusselt number $\Nu$ \eqref{eq-Nu} and the final value of the instantaneous Nusselt number $\Nu(\mathcal T)$, versus iteration, normalised by values for the unforced case. 
Changes in these quantities
show good agreement, indicating that the  global (volume integrated) scalar dissipation quantity effectively captures the local (boundary) heat transfer behaviour measured by the Nusselt number.  
Figure \ref{prelical}$(c)$  shows the residual of the calculation, which drops quickly in first $50$ iterations then more gradually.
(Spikes are related to the method that seeks to increase $\epsilon_0$ as much a possible, which affects the magnitude of the residual via $\lambda_0$.) 
Usually, the optimisation is stopped when the change in the Nusselt number drops below $10^{-5}$.  
Figure \ref{prelical}$(d)$ tracks the amplitude of the three components of the body force versus iteration. The amplitude of the streamwise component drops significantly, while the amplitude of the cross-stream components increases. This suggests that the cross-stream components of the body force play a dominant role in enhancing heat transfer, whereas the contribution of the streamwise component is nearly negligible. 

\begin{figure}
         	\centering
         	\includegraphics[angle=0,width=1\textwidth]{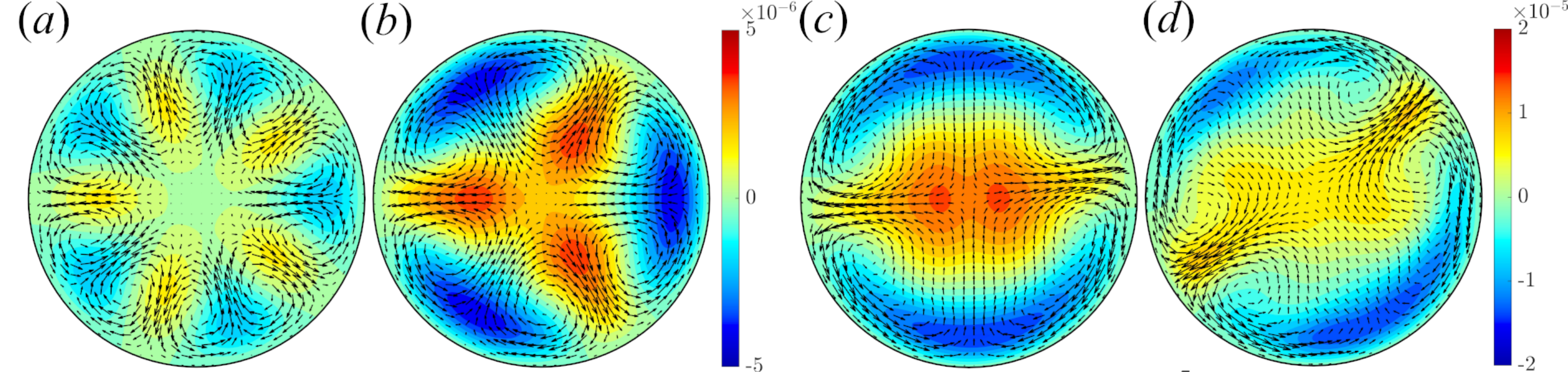}  
         	\caption{Optimised body forces for target time horizons $(\textit a)$ $\mathcal T=50, (\textit b)$ $\mathcal T=100, (c)$ $\mathcal T=200, (\textit d)$ $\mathcal T=600$ at $C=3, A_0=5e-7, Re=3000, N=1$. Colours indicate the streamwise component, while
           arrows represent the cross-stream components of the body force. 
           $(\textit {a})$ and 
           $(\textit {b})$ share the first colourbar, 
           $(\textit {c})$ and 
           $(\textit {d})$ share the second. 
           The largest arrow has magnitude $2.38 \times 10^{-4}$ in $(\textit a)$, $2.36 \times 10^{-4}$ in $(\textit b)$, $2.93 \times 10^{-4}$ in $(\textit c)$ and $3.21 \times 10^{-4}$ in $(\textit d)$.}  
         	\label{Force-T}
        \end{figure}

   If the time horizon $\mathcal{T}$ is long enough, then $\Nu$ is optimised for the 
steady response to the force. Figure \ref{prelical}(\textit {a}), suggests that $\mathcal{T}$ should be greater than $200$.  Indeed, the form of the optimal is found to change when increasing $\mathcal{T}$ from $50$ to $100$, and again to $200$, but increasing further to $600$, the optimal is essentially the same up to a rotation, as shown in figure \ref{Force-T}.
    Interestingly, the optimal body force optimised for a short target time has perfect rotational symmetry. 
\AW{That the force induces rolls is consistent with the steady velocity fields computed by \cite{meng2005field}.
\EM{The time horizon}    $\mathcal{T}$ is an extra parameter here}, and as it is  
    increased, the azimuthal wave number $m$ decreases. This is consistent with smaller-scale vortices growing more rapidly \citep{schmid2007nonmodal}, thereby increasing the heat transfer within a shorter time.
With a longer target time, however, the  larger-scale mode is more effectively amplified for the given magnitude of force. 

       It is found that the distribution of the body force is almost uniform in the streamwise direction.
       \AW{
Although streamwise dependence of the force was included as a possibility in the preliminary calculations, it was not expected -- structures sweep rapidly through the domain, leading to streamwise independence in the time-integrated update to $\vec{F}^{(j)}$ in the optimisation.  In the isothermal calculations of \cite{marensi2020designing},
streamwise-dependent forces were not identified even with domains 5 times longer.}
Therefore, to simplify the form of the body force and 
   to accelerate convergence,
   we constrain the body force to be streamwise-independent 
 in the optimisations of the following sections
   by zeroing coefficients of the streamwise-dependent Fourier modes. 

  \subsection{Optimisation in the laminarisation regime} 
  \label{sec:laminarisation regime} 
   \subsubsection{Optimal force for the laminar state}
    Having examined properties of the parameters necessary for optimisation, in this section we 
    consider optimisation in the laminarised regime at $C=3,\Rey=3000$ in more detail. In particular, we examine the dependence of the rotational symmetry on $A_0$ and the presence of local optimals (dependent on the initial guess for the force).

\begin{figure}
         	\centering
         	\includegraphics[angle=0,width=0.6\textwidth]{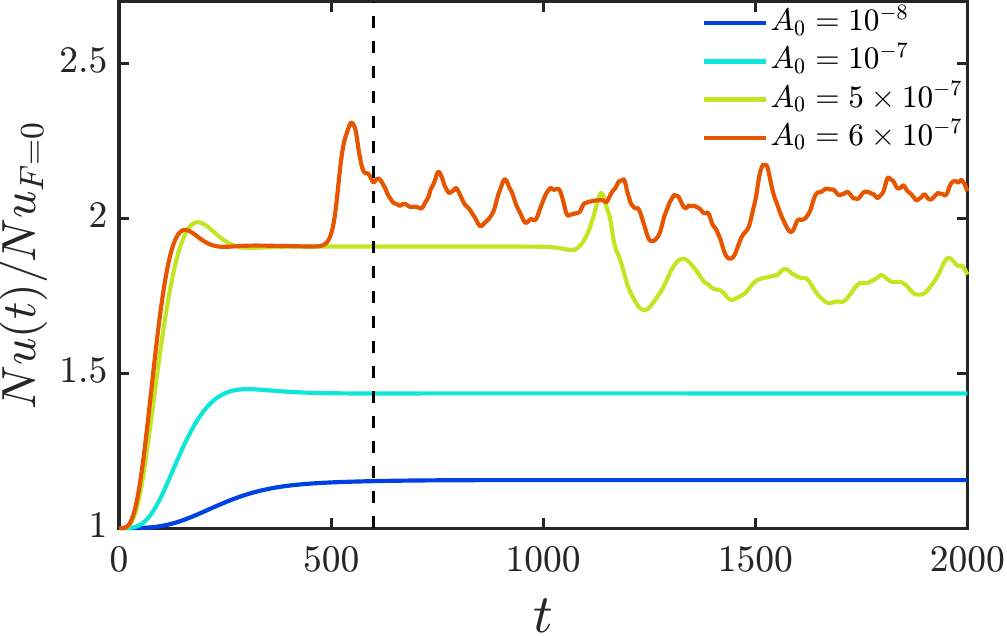}  
         	\caption{ Instantaneous Nusselt number normalised by the value for unforced flow
          for several force amplitudes, starting from a
          laminar initial condition at $C=3, Re=3000$. The vertical dashed line indicates the optimisation target time $\mathcal T=600$.
          }    
         	\label{Nu-A0}
    \end{figure}

For the laminar initial condition, as we have assumed streamwise independence for the force,
a small streamwise-dependent perturbation 
must be added to the initial velocity 
so that transition to turbulence may be triggered if the resulting two-dimensional flow is unstable.  We add a perturbation of  magnitude $E_0=\frac{1}{2}\langle \vec{u}^2\rangle \approx 10^{-7}$
and    
    set a longer $\mathcal{T}=600$ to allow the occurrence of transition.
Figure \ref{Nu-A0} shows the instantaneous Nusselt number, $\Nu(t)$, for 
several $A_0$.
 Optimisation improves the heat transfer substantially: 
for $A_0=10^{-7}$ heat transfer is almost 
50\% greater than that of the unforced flow,
and for $A_0=5\times 10^{-7}$ is almost doubled. For slightly higher $A_0=6\times 10^{-7}$, $\Nu(t)$ experiences a sudden increase near the end of the optimisation target time and fluctuates thereafter, indicating the onset of  turbulence. 
(At $A_0=5\times10^{-7}$, transition is observed very late, at around $t=1000$, and  interestingly, the transition does not lead to a larger $\Nu$. This phenomenon will be discussed later.) 
        
\begin{figure}
         	\centering
         	\includegraphics[angle=0,width=0.9\textwidth]{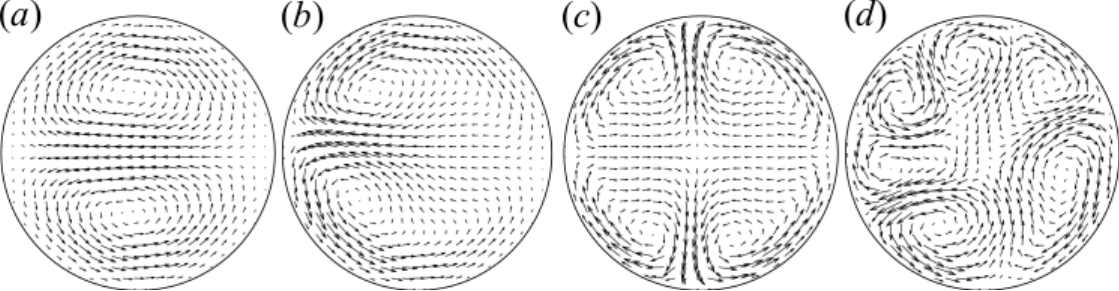}  
         	\caption{ \EM{Optimised body forces} at $(\textit a)A_0=10^{-8}, (\textit b)A_0=10^{-7}, (\textit c)A_0=5\times10^{-7}, (\textit c)A_0=6\times10^{-7}$, starting from a laminar initial condition at $C=3, \mathcal T=600, Re=3000$. 
            The largest arrow has magnitude $4.85 \times 10^{-5}$ in $(\textit a)$, $1.87 \times 10^{-4}$ in $(\textit b)$, $4.05 \times 10^{-4}$ in $(\textit c)$ and $4.26 \times 10^{-4}$ in $(\textit d)$. 
         As the axial component is at least an order of magnitude smaller, it is not shown.
          }          
         	\label{OF-lam}
        \end{figure}

    The typical amplitude-dependent form of the forces obtained from optimisations are given in figure \ref{OF-lam}(\textit{a-d}). 
(As the axial component is at least an order of magnitude smaller, it is not shown.)
    At small $A_0$, the body force has a single pair of rolls. 
    At increased force amplitude,
    figure \ref{OF-lam}({\it b})
    illustrates how the vortex structure 
    gradually approaches the wall, reducing the spatial scale in both the radial and  spanwise directions.  This is actually found to be a local optimal
    for this $A_0$, as two pairs of rolls may be squeezed in to increase $\Nu$ a little further.
    At larger $A_0$, more rolls are seen in     
    figure \ref{OF-lam}(\textit{c-d}).  
    For the largest $A_0$, turbulence is triggered within $\mathcal{T}$, and the optimisation struggles to converge to a well-organised optimal force.  However, a preference for roll structures of larger $m$ is clear.
    The form and increase in wavenumber is consistent with the calculations of  \cite{meng2005field,jia2014convective, wang2015application}
    for steady flow.

We have observed that optimal body forces display rotational symmetry of different azimuthal wave numbers $m$ in figures \ref{Force-T} and \ref{OF-lam}.
We let $O_2$ denote an optimal with $2$-fold rotational symmetry.  This optimal may have non-zero Fourier coefficients only for $m=0,2,4,6,8,...$, but note that this does not exclude a force with only non-zero modes $m=0,4,8,...$, which corresponds to an optimal $O_4$ with $4$-fold rotational symmetry. 
To examine the influence of the azimuthal periodicity,  and to simplify the optimisation further, we consider optimal forces restricted to azimuthal Fourier modes of wavenumbers $0$ and $m$ only, and denote them $O_{Fm}$.  
Note that rotational symmetry is imposed only on the force, and not on the velocity field.

\begin{figure}
         	\centering
         	\includegraphics[angle=0,width=1 \textwidth]{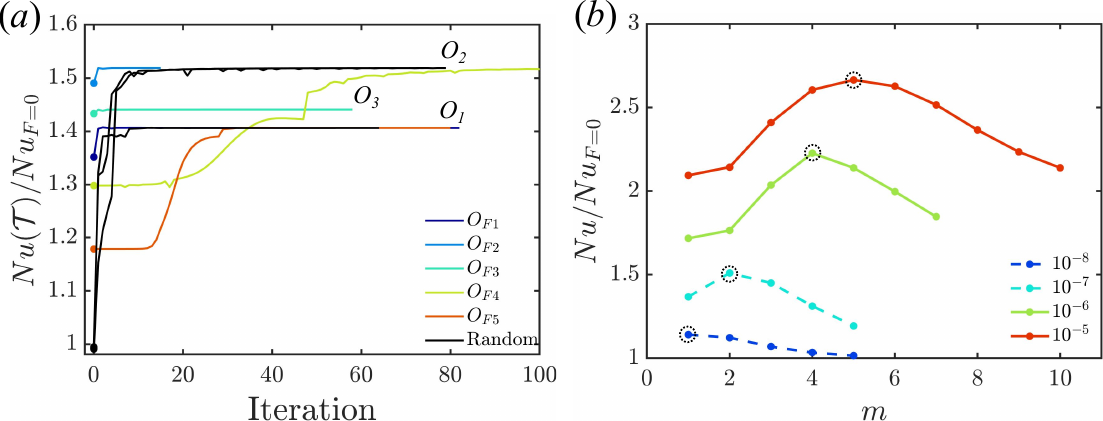}  
         	\caption{ 
            (\textit{a}) Nusselt number of final state $\Nu(\mathcal{T})$
            versus iteration for different initial forces, with $A_0=10^{-7}, C=3, Re=3000$.  The legend indicates the initial force, and the resulting local optimal forces are labelled on the curves. (\textit{b}) Normalised (time-averaged) Nusselt numbers at $C=3, Re=3000$ versus $O_{Fm}$ for force amplitudes $A_0$ indicated in the legend. 
            The global optimal is highlighted with a dashed circle.  
            For dashed lines, the forced flow state remains laminar. For solid lines, the forced state is turbulent and the values are time-averages.
          }         
         	\label{Nu-LOP}
        \end{figure}

Optimisations have been computed for $O_{Fm}$, then used as starting forces for optimisations in the full space $O_1$.  In this way, we examine the dependence on $m$ to determine which 
rotational symmetry is the global optimal.
The Nusselt numbers of the final states, $\Nu(\mathcal{T})$, as a function of iteration are shown in figure \ref{Nu-LOP}($\textit {a}$) starting from the $O_{Fm}$ forces. Three optimisations starting from random initial forces are also shown. 
Several observations can be made.  Firstly, 
there are multiple local optimals $O_m$, of which $O_2$  (figure \ref{OF-lam}(\textit{c})) is the global optimal
for this $A_0$. 
Secondly, the optimal of type $O_m$ does not produce much greater $\Nu$ than the optimal $O_{Fm}$, of the reduced Fourier space, used as the starting force.
Thirdly, if the starting force is quite perturbed, such as for the random initial forces, it is most likely to end up at the global optimal.  Similarly, the optimisation starting from \CSJ{$O_{F4}$ (light green line)} appears to pass close to $O_1$, but ends at the global optimal $O_2$.

    As we have observed that $O_m$ does not produce much greater $\Nu$ than $O_{Fm}$, 
  for small $A_0$ at least,
    we directly compare $\Nu$ of the flow forced by $O_{Fm}$ for several $m$ to determine the global optimal.  
    Figure \ref{Nu-LOP}(\textit{b}) shows the Nusselt number, calculated using averages over $5000$ time units for each simulation.
    For the dashed lines, 
    the final state is still laminar
    and good convergence is easily achieved.
    For the solid lines at larger $A_0$, the final state is turbulent, which 
    renders convergence difficult.
    For the latter case, and for convenience in this section, the force has been 
    calculated using an artificially stabilised two-dimensional flow,
    by putting $K=1$ in (\ref{eq:discretisation}) (equivalent to adding no three-dimensional perturbation, $E_0=0$). The optimal force leads to an artificially stabilised optimal velocity field, 
    as for the optimal steady velocity fields reported by  \cite{meng2005field,jia2014convective,wang2015application} .
    The force is then applied, here resulting in a fully three-dimensional time-dependent turbulent simulation,
    from which $\Nu$ is calculated.
    (In principle, the 
    stabilisation of the flow during optimisation may render the force no longer optimal. This is examined further
    for each initial flow regime, and is found to be a good approach for
    intermediate $C$.  We will show this for the convective case in \S \ref{section:convective turbulence}.)  %
    As $A_0$ increases, the rotational symmetry $m$ of the (global) optimal increases, from $m=1$ at $A_0=10^{-8}$, to $m=2$ at $A_0=10^{-7}$, and is $m\approx 5$ at substantially larger $A_0=10^{-5}$.  However, it should be noted that $\Nu$ is not strongly dependent on $m$.

\subsubsection{
The path to transition and effect on heat transfer}

\begin{figure}
         	\centering
         	\includegraphics[angle=0,width=1 \textwidth]{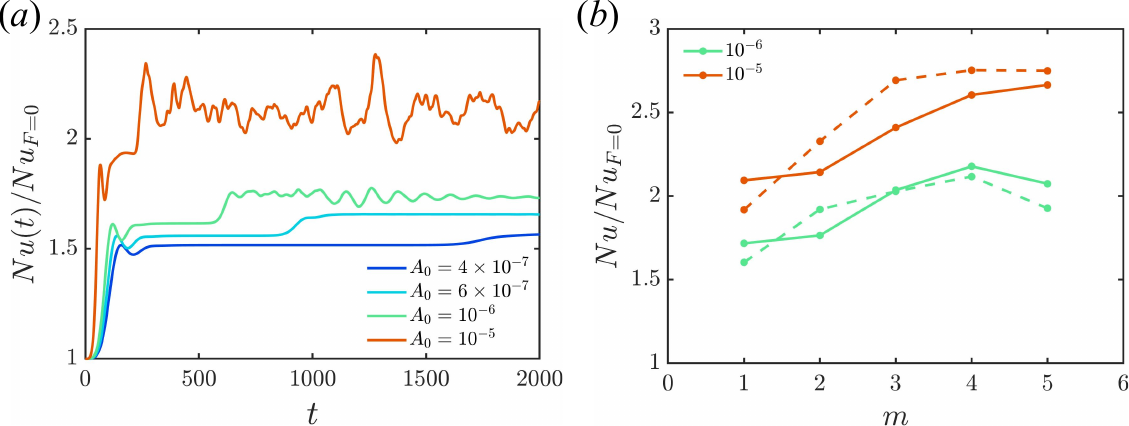} 
         	\caption{ 
           Enhanced $\Nu$ of forced flows at $C=3, Re=3000$. 
           $(\textit a)$ 
           Instantaneous $\Nu(t)$ for flows forced by $O_{F1}$. $(\textit b)$  $\Nu$ (time-averaged) for flows forced by $O_{Fm}$
           in the turbulent state (solid lines) and 
           $\Nu(\mathcal{T})$ for
           the 
           stabilised streamwise-independent state (dashed lines).}   
         	\label{Nu-t-m}
        \end{figure}

\begin{figure}
         	\centering
         	\includegraphics[angle=0,width=1 \textwidth]{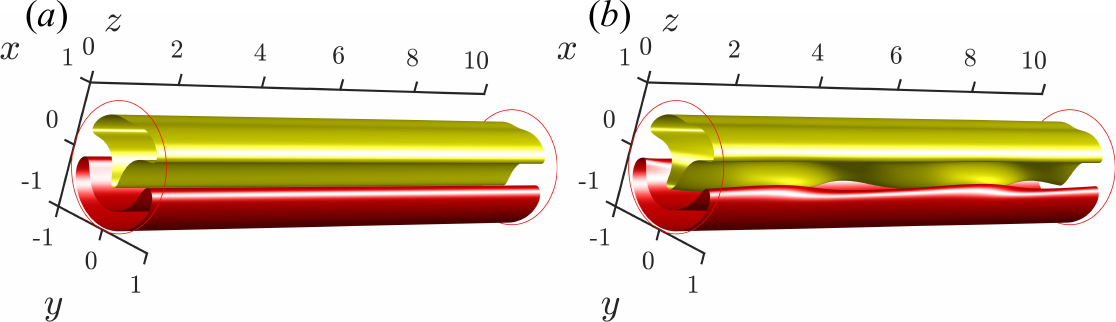}  
         	\caption{ Isosurfaces of streamwise velocity for $(\textit a)$ the streamwise-independent state at $t=500$ and $(\textit b)$ the travelling wave solution at $t=1500$, for flow forced by $O_{F1}$ with $A_0=6\times 10^{-7}$. Red/yellow are at 20\% of the min/max streamwise velocity.
          }        
         	\label{isoTW}
        \end{figure}

     Figure \ref{Nu-t-m}(\textit{a}) shows the instantaneous Nusselt number $\Nu(t)$ for flows forced by $O_{F1}$ at several force amplitudes.   At small $A_0$, the flow is reshaped into a two-dimensional forced laminar state.
With an increase of force amplitude to $A_0=6\times 10 ^{-7}$, the flow does not transition to turbulence directly --- instead, the two-dimensional state quickly forms, then transitions later to a travelling wave solution.
 Figure \ref{isoTW} shows the two-dimensional reshaped laminar solution and a travelling wave solution found in the flow forced by $O_{F1}$ at $A_0=6\times 10^{-7}$
($t\approx 500$ and $t\approx 1500$ of figure \ref{Nu-t-m}(\textit{a}) respectively). 
In isothermal flow, forces have been used to find travelling wave solutions, via homotopy
\EM{\citep{faisst2003traveling,wedin2004exact}}, 
but were
only found by this method for higher rotational symmetry $m\ge 2$.
The travelling wave solution has larger $\Nu$ compared with the two-dimensional reshaped laminar solution. With further increase of force amplitude, the flow transitions 
from the two-dimensional state to a (mildly) chaotic three-dimensional state, and finally also converges to a travelling wave state at a later time (not shown here). At $A_0=10^{-5}$, the flow directly transitions to a strong chaotic three-dimensional state, along with a greatly increased $\Nu$.
However, this is not always the case, and in fact figure \ref{Nu-t-m}(\textit{b}) shows that at larger $A_0$ the transition
from two-dimensional flow
(stabilised by setting $K=1$)
to the chaotic
three-dimensional state \EM{leads to} a decrease in $\Nu$, \AW{except for the case $m=1$}. 
         \begin{figure}
         	\centering
         	\includegraphics[angle=0,width=1 \textwidth]{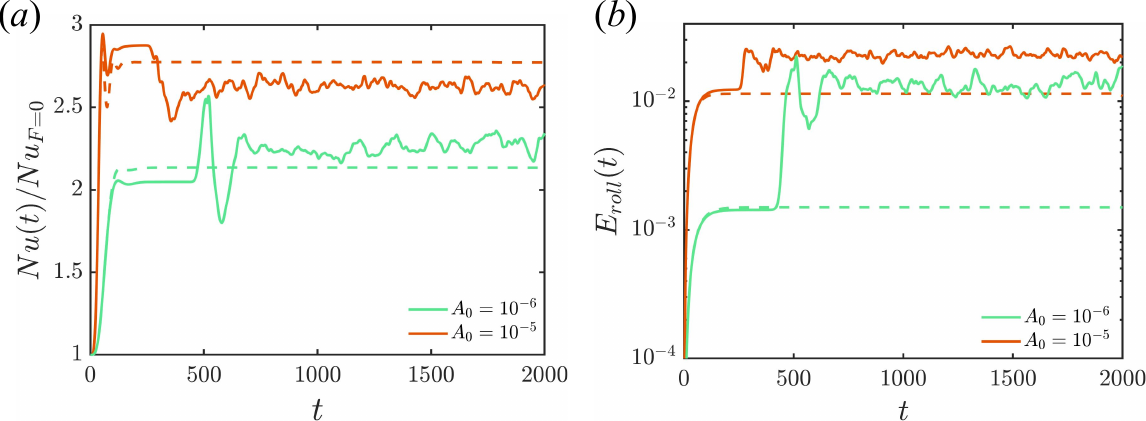} 
         	\caption{Time series of (\textit{a}) instantaneous $\Nu(t)$ and (\textit{b}) instantaneous energy of rolls $E_{roll}=E(u_r)+E(u_\phi)$, $E(u_r)=\frac{1}{2}\langle u_r^2 \rangle$, $E(u_\phi)=\frac{1}{2}\langle u_\phi^2 \rangle$, for flow forced by $O_{F4}$ at $C=3, Re=3000$ at two different $A_0$. 
            Solid lines are for the forced turbulent state, and dashed lines are for the  streamwise-independent (artificially stabilised) state.
            }
         	\label{Er-m23}
        \end{figure}

\begin{figure}
         	\centering
         	\includegraphics[angle=0,width=1 \textwidth]{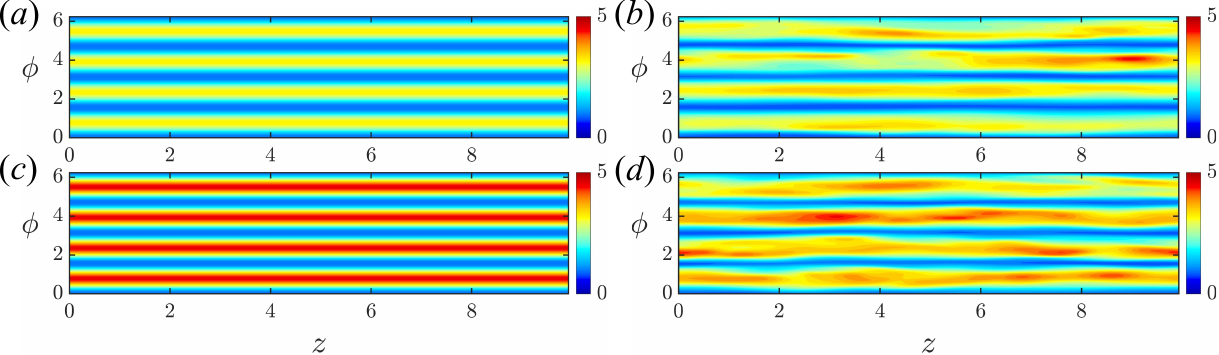} 
         	\caption{ Contours of the radial temperature gradient $ \frac{\partial T}{\partial r}$ at the boundary at $C=3, Re=3000$, normalised by the unforced laminar radial temperature gradient. The flow is forced by $O_{F4}$ with $(\textit{a,b})$ $A_0=10 ^{-6}$ and $(\textit{c,d})$ $A_0=10 ^{-5}$, the flow states are taken in $(\textit{a,c})$ from the forced laminar state ($t=1000$) and $(\textit{b,d})$ forced turbulent state ($t=1000$).}     
         	\label{Contour-nu-m23}
        \end{figure}

It is interesting that the more chaotic 
state does not  
necessarily lead to an improvement in heat transfer. 
    For the case where the flow is forced by $O_{F4}$, the instantaneous Nusselt number $\Nu(t)$ and the roll energy $E_{roll}(t)=E(u_r)+E(u_\phi)$  are shown in figure \ref{Er-m23} for two amplitudes $A_0$.
    At $A_0=10^{-6}$
    heat transfer increases after transition, but at $A_0=10^{-5}$ it is reduced after transition.
In both cases, the energy of rolls increases, but in the latter case by not as much.
Stronger rolls are typically associated with enhanced heat transfer, but this shows that higher roll energy alone does not necessarily correspond to more efficient heat transfer. 
Figure \ref{Contour-nu-m23} shows the contours of the radial temperature gradient ${\partial T}/{\partial r}$ evaluated at the boundary, for the forced laminar and forced turbulent states,
 normalised by the mean radial temperature gradient of the unforced flow.
    When the flow is forced by $O_{F4}$ with $A_0=10^{-6}$, the laminar state (figure \ref{Contour-nu-m23}(a)) exhibits distinct regions of strong and weak heat transfer. After the transition (figure \ref{Contour-nu-m23}(b)), these regions are not fixed,
   and the regions of higher heat flux widen and intensify a little.
    At $A_0=10^{-5}$ (figure \ref{Contour-nu-m23}(c-d)), the regions of higher heat flux also widen after the transition, but are less organised and weaker than for the steady flow, despite the slightly increased roll energy.
This suggests that the streamwise vortices in the forced turbulent states are not as efficient as those in the forced laminar states. 
 The rolls are unsteady in forced turbulent states and move further from the wall intermittently due to the waving of low-speed streaks, leading to a nonuniform heat transfer distribution.  Heat transfer is enhanced on the sides where the rolls are positioned close to the wall, but weakens as the rolls move further away from the wall.     
 Overall, the transition to the forced turbulent state involves two main competing effects on heat transfer: the first is the enhancement by the rolls, which increases heat transfer by facilitating better mixing, the second is unsteadness of the rolls, which at larger $A_0$ can reduce heat transfer due to inconsistency in their position near the wall. 
The case $O_{F1}$ is an exception
in figure \ref{Nu-t-m}(\textit{b}),
where the transition to a chaotic state leads to an increase in $\Nu$ 
-- only one pair of rolls is inefficient and the unsteadiness can lead to the creation of additional rolls.

\EM{In this section, we have shown that optimisations in the laminarised regime can substantially enhance heat transfer, with the optimal forcing inducing axially aligned rolls of increasing azimuthal symmetry — though with only a weak dependence of $Nu$ on $m$. Interestingly, triggering turbulence does not necessarily lead to an increase in $\Nu$, as the resulting unsteady rolls are less efficiently positioned for heat transfer.}

  \subsection{Optimisation in the shear turbulence regime} 
  Returning to smaller $C$, the shear-turbulence case is the most challenging, as the flow state remains highly chaotic.  As turbulence is already effective in enhancing heat transfer relative to the laminarised case, it is not obvious that optimisation should be able to improve heat transfer substantially.

We focus on the case $C=1$ at $\Rey=3000$.
For the highly chaotic flow, it is difficult to apply the method with a large target time
\citep{pringle2010using,pringle2012minimal,marensi2019stabilisation},
and convergence was found to fail for $\mathcal{T}$ even as low as $100$.  However, reasonably good convergence was found for $\mathcal{T}=50$.  Although this is not sufficient time to capture the statistics of the end state, it is sufficient time for a response to be observed, so that it is reasonable to examine whether the heat transfer has been pushed in the right direction.

\begin{figure}
         	\centering
         	\includegraphics[angle=0,width=0.6 \textwidth]{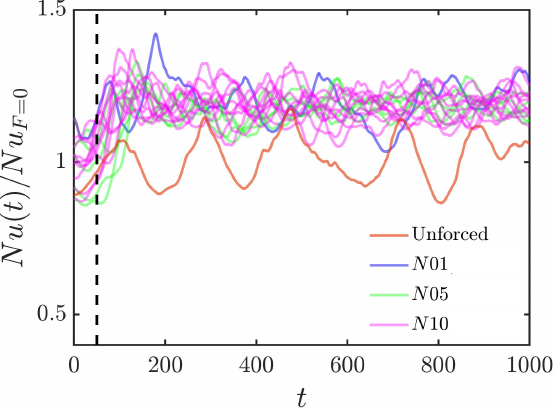}  
         	\caption{Instantaneous Nusselt number for flows forced by  $O_1$ at $A_0=10^{-6}$,
            where the legend indicates the number of initial velocity fields used in the optimisation.
            The time horizon $\mathcal{T} =50$ is marked by the vertical dashed line. 
              }                 
         	\label{OF-T-C01}
        \end{figure}

\begin{figure}
         	\centering
         	\includegraphics[angle=0,width=0.75 \textwidth]{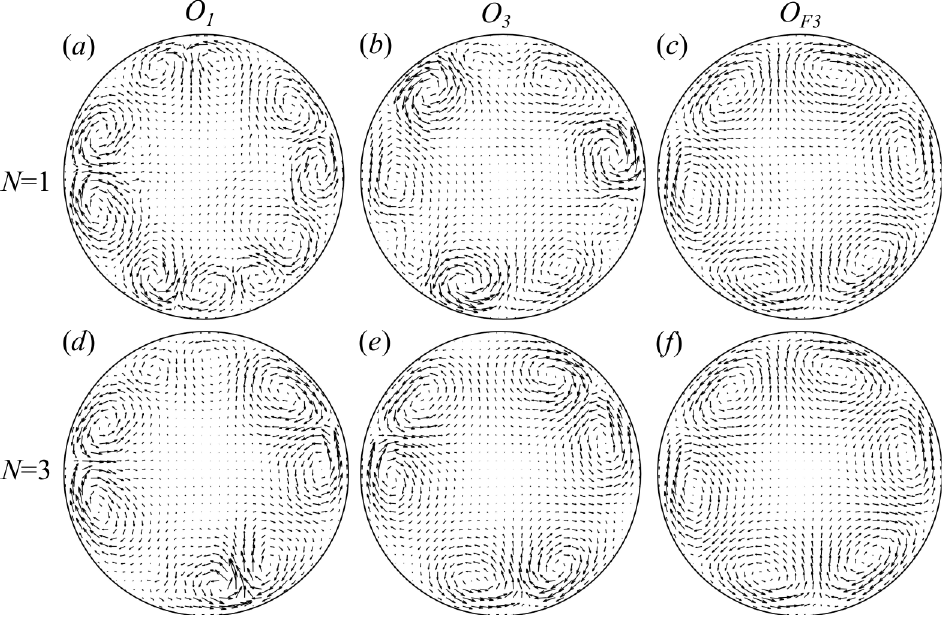}  
         	\caption{ 
             Forces \EM{$O_1$, $O_3$ and $O_{F3}$}
            optimised using $N$ initial velocity fields at $A_0=10^{-6}$, $C=1$. 
             The largest arrow has magnitude $6.95\times 10^{-4}$ in $(\textit {a})$, $7.52\times 10^{-4}$ in $(\textit {b})$, $5.76\times 10^{-4}$ in $(\textit {c})$, $8.14 \times 10^{-4}$ in $(\textit {d})$, $6.60\times 10^{-4}$ in $(\textit {e})$, $5.76\times 10^{-4}$  in $(\textit {f})$.
            }                 
         	\label{OF-C01-IC}
        \end{figure}

Figure \ref{OF-T-C01}$(a)$ shows the instantaneous Nusselt number $\Nu(t)$ for flows forced by the full space optimal $O_1$
with $A_0=10^{-6}$.
The optimisation, although with a short target time, still increases the Nusselt number significantly. 
Surprisingly,
table \ref{tab:NuIC} shows that the (subsequent time-averaged) $\Nu/\Nu_{F=0}$ 
does not change significantly with more initial conditions $N$, despite that the short $\mathcal{T}$ might suggest greater dependence on the initial condition, nor does $\Nu$ vary significantly with rotational symmetry.
Structures of the optimised forces are shown in figure \ref{OF-C01-IC} \EM{for $N=1$ and $3$ (the cases for $N=9$ are almost identical to $N=3$)}. Although $\Nu$ varies very little with $N$, for $O_1$ the structure of the optimal force does change --
going from $N=1$ to \EM{$3$}
initial conditions, the force develops towards a structure more like the $O_3$ optimal, but larger $N$ 
slightly
improves convergence.  Optimising for $O_3$ itself, the structure clearly becomes more regular as $N$ is increased.
For $O_{F3}$ \EM{(third column)}, $O_{5}$ 
and $O_{F5}$ \EM{(not shown)}, 
the structure of the force does not change for larger $N$, 
so that only one initial velocity field is sufficient.
This is reasonable for larger $m$, 
as the angular section $[0,2\pi/m]$ of the force is determined by $m$ angular subsections of the rotationally unconstrained velocity field.

\begin{figure}
         	\centering
         	\includegraphics[angle=0,width=1 \textwidth]{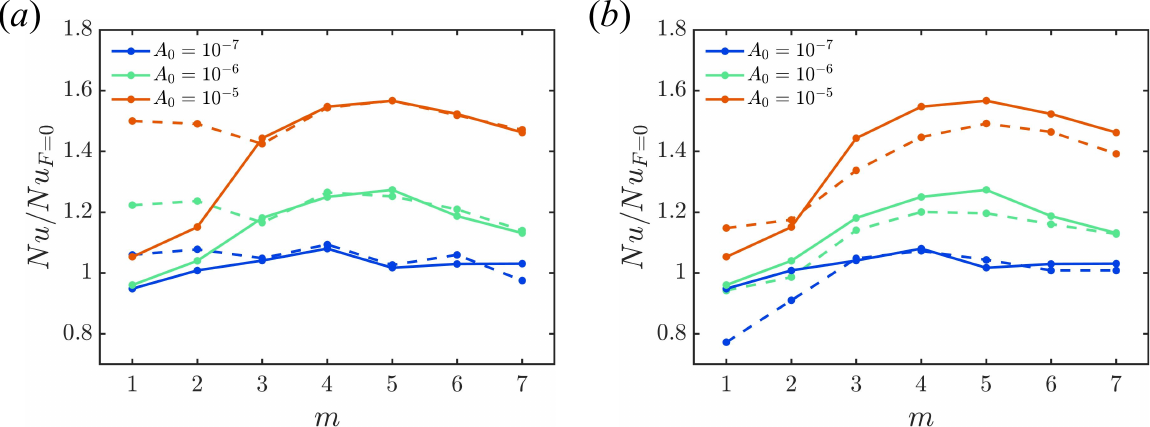} 
         	\caption{ Normalised Nusselt numbers for flows at  $C=1, Re=3000$ subject to optimal forces in the full and reduced rotational symmetries.  $(\textit a)$ Comparison between $O_{Fm}$ (solid lines) and $O_m$ (dash lines). 
            $(\textit b)$ Comparison between $O_{Fm}$ optimised for shear turbulence (solid line) and a steady two-dimensional laminar state (dashed line).
            }
         	\label{Nu-A0-m-C01}
        \end{figure}

\begin{figure}
         	\centering
         	\includegraphics[angle=0,width=0.8 \textwidth]{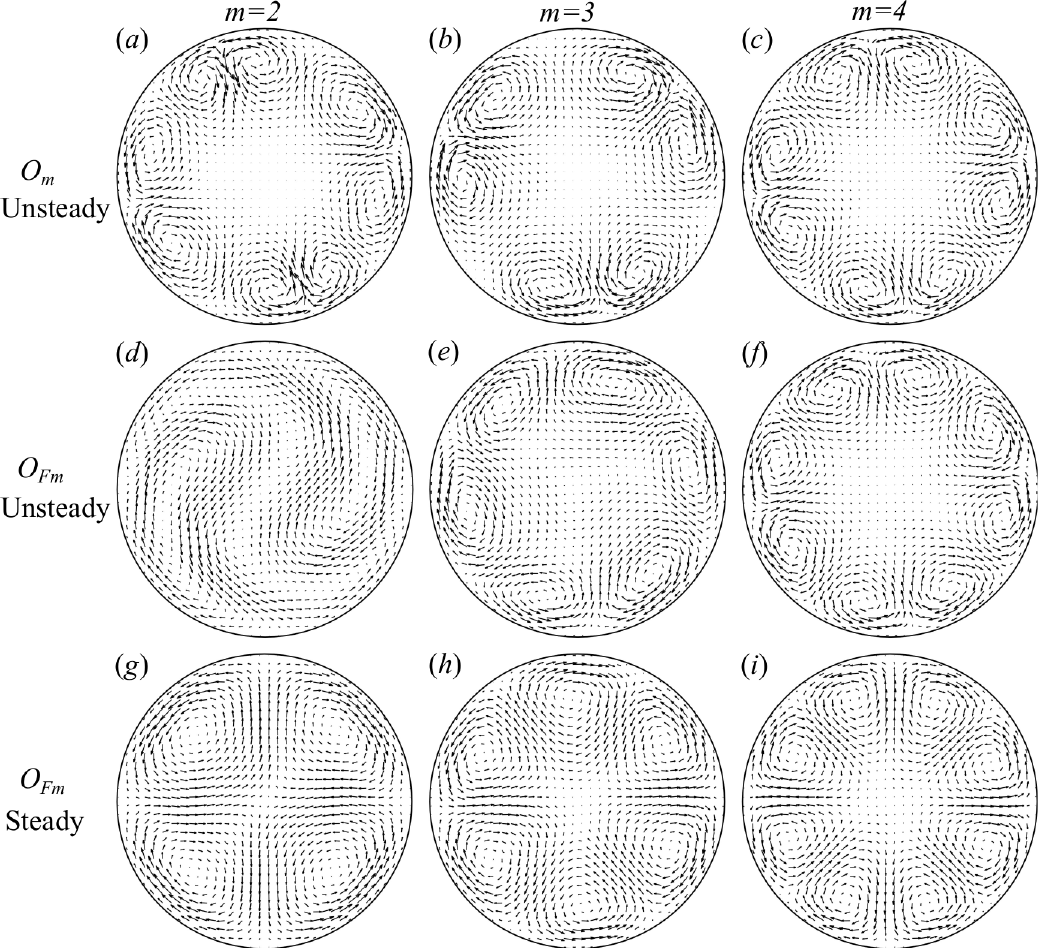} 
         	\caption{
            Optimal forces at $A_0=10^{-6}, C=1$.
The largest arrow has magnitude $6.01 \times 10^{-4}$ in $(\textit {a})$, $6.60\times 10^{-4}$ in $(\textit {b})$, $5.50\times 10^{-4}$ in $(\textit {c})$,$5.15\times 10^{-4}$ in $(\textit {d})$, $5.76\times 10^{-4}$ in $(\textit {e})$, $5.40\times 10^{-4}$ in $(\textit {f})$, $6.34 \times 10^{-4}$ in $(\textit {g})$, $5.55\times 10^{-4}$ in $(\textit {h})$, $5.99\times 10^{-4}$ in $(\textit {i})$. $N=1$ initial velocity condition is used for $O_{Fm}$ and $N=3$ initial velocity conditions are used for $O_{m}$. 
            \AW{In the `steady' case, axial-independence is imposed on the velocity which results in a steady response velocity to the force.}
        }         
         	\label{OF-m-C01}
        \end{figure}

Although we have computed optimals only for a short $\mathcal{T}$, 
our results strongly suggest that inducing rolls remains optimal even for flow already
turbulent, in the shear-turbulence regime.  As before, it is interesting to examine whether or not there is a strong dependence on the rotational symmetry $m$. 
       \begin{table}
    \begin{center}
    \def~{\hphantom{0}}  
     \begin{tabular}{lccccl}
      $N$& $O_{1}$& $O_{3}$& $O_{5}$& $O_{F3}$&$O_{F5}$\\
      1& 1.22& 1.16& 1.25& 1.18&1.27\\
      3& 1.22& 1.18& 1.27& 1.18&1.26\\
      9& 1.20& 1.18&1.27&1.17&1.25\\
     \end{tabular}
     \caption{$\Nu/\Nu_{F=0}$ for flows  forced by optimal forces indicated. $N$ is the number of initial velocity conditions used in the optimisation. }
     \label{tab:NuIC}
      \end{center}
  \end{table}
%
Figure \ref{Nu-A0-m-C01}($\textit {a} $) shows the Nusselt number for flows forced by $O_{Fm}$ (solid) and $O_{m}$ (dashed)  for several $m$,
using $N=1$ for $O_{Fm}$ and $N=3$ for $O_m$.
For small $m=1,2$, the force $O_m$ appears to produce larger $\Nu$ than $O_{Fm}$, but this is because the optimisation for $O_1$ actually found a structure closer to that for $O_3$, previously seen in figure \ref{OF-C01-IC} (cf. \EM{(e) and (f)}),
and similarly, the $O_2$ optimal is structurally more like $O_4$,
seen in figure \ref{OF-m-C01} (cf. (a) and (c)).
Constraining the number of rolls strictly, by using the single Fourier mode, the rolls of $O_{F3}$ are a little stretched (figure \ref{OF-m-C01}\EM{(\textit{e})})
relative to those of $O_3$ (figure \ref{OF-m-C01}(\textit{b})), but for $m>3$ there is essentially no visible difference between $O_m$ and $O_{Fm}$.  For $m\ge 3$, $\Nu$ is relatively insensitive to the wavenumber (figure \ref{Nu-A0-m-C01}($\textit {a}$)).

An interesting observation is the occurrence of laminarisation of the shear turbulence 
when forced by $O_{F1}$ at low amplitudes $A_0=10^{-7}$. 
A very similar observation was reported by \cite{willis2010optimally}, where there the roll-force was beneficial for drag reduction in isothermal flow.  Here, as the turbulent flow enhances heat transfer, the laminarisation can lead to a reduction in the heat transfer, i.e.
$\Nu/\Nu_{F=0}<1$, see figure \ref{Nu-A0-m-C01}($\textit {b}$).

Optimisation for a steady laminar flow can be imposed by setting $K=1$ in (\ref{eq:discretisation}), as applied in \S \ref{sec:laminarisation regime}. 
Figure \ref{Nu-A0-m-C01}($\textit {b} $)
compares optimisation with the short 
$\mathcal{T}$ (solid lines) with  
steady laminar optimisation (dashed lines) with $\mathcal{T}=600$.
Particularly for larger $A_0$, it should be noted that including the time dependence of the flow in the optimisation does improve the resulting $\Nu$ over the steady assumption, 
despite the short $\mathcal{T}$.
There is an exception for the $m=1,2$ case at the largest $A_0$, however, where the short-time optimal results in an unusual force structure, see, e.g., figure \ref{OF-m-C01}(\EM{\textit{e}}).
For the structure of the optimal forces,
when optimised for the steady two-dimensionalised state (figure \ref{OF-m-C01}(\EM{\textit{g-i})}) these forces may induce flows that are similar to those of previous calculations
\citep{meng2005field,jia2014convective,wang2015application}, although there the optimisations were independent of the flow regime.

\AW{Overall, it is found that for the 
shear-turbulent regime, rolls remain optimal, but the best choice of azimuthal wavenumber may differ from that in the laminarised flow regime, i.e.\ there is dependence on $C$ through the governing equation\EM{s}.  Including time-dependence in the optimisation,
the rolls are notably closer to the wall
(figure \ref{OF-m-C01}(\textit{d-f})). 
It is possible that this difference is linked to the flattened turbulent mean velocity profile in the shear-turbulence state, which leads to a lift-up process more localised towards the near-wall region.
}


  \subsection{Optimisation in the convective turbulence regime} 
  \label{section:convective turbulence}
    Here, optimisation was first considered for a  weakly convective turbulent state at  $C=4$.
    As the velocity fields in the convective state are time-dependent, optimisations were initially performed using several initial velocity fields 
    ($N>1$) at this $C$ and random initial forces.
However, even for small $A_0=10^{-7}$ it was found that the flow is rapidly laminarised by the force, 
\AW{it approaches a two-dimensional streamwise-independent state with enhanced $\Nu$.  Therefore in this case,} 
like at $C=3$ for the laminarisation regime, $N=1$ is sufficient.
   Also like the laminarised case, 
   the optimal Nusselt number shows substantial improvement compared to the unforced case, 50\% at $A_0=10^{-7}$,
   and the structure of the optimal,
   although calculated for $O_1$, looks 
   very similar to that of figure 
   \ref{OF-lam}({\it c}), close to $O_{2}$ symmetry. 
Due to the laminarisation by the new force at $C=4$, optimisations for $O_{Fm}$ exhibit similar behaviour to those for the laminarised case at $C=3$.   

        \begin{figure}
         	\centering
         	\includegraphics[angle=0,width=0.5 \textwidth]{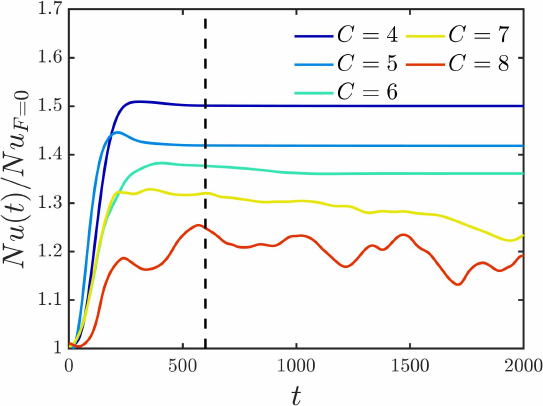}  
         	\caption{
            Flows forced by $O_1$ stabilised to larger $C$.
            Here $A_0=10^{-7}$. The vertical dashed line indicates the optimisation target time $\mathcal T=600$.}     
         	\label{Nu-A0-C04}
        \end{figure}

     We therefore optimise heat transfer at larger $C$ to examine how far this laminarisation phenomenon occurs.  Time evolutions for $\Nu(t)$  optimised at $C=4-8, \mathcal T=600, A_0=10^{-7}, 
    $ for $O_1$ are shown in figure \ref{Nu-A0-C04}.   With increased buoyancy force, laminarisation by the force still occurs at $C=5,6$, but disappears at $C=7, 8$.  Although the turbulence does not decay, the Nusselt number at $C=7$ only fluctuates slightly.  
   At $C=8$, the amplitude of fluctuations in $\Nu(t)$ are similar to those of the unforced flow, but with a higher mean value.
   
\begin{figure}
         	\centering
         	\includegraphics[angle=0,width=1 \textwidth]{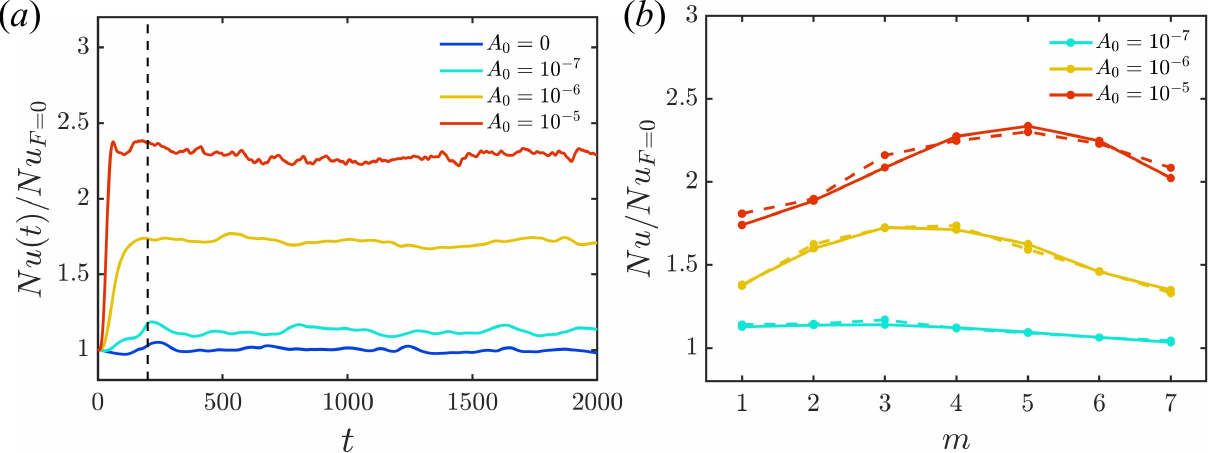}  
         	\caption{
            (\textit{a}) Instantaneous Nusselt number for flow forced by $O_{F4}$ at several $A_0$ at $Re=3000, C=8$. The vertical dashed line indicates the optimisation target time $\mathcal T=200$. (\textit{b}) Comparison between $O_{Fm}$ optimised with unsteady convective turbulence (solid line) and with a two-dimensional steady velocity (dash line).}     
         	\label{Nu-A0-C08}
        \end{figure}

   Due to the laminarisation, the optimisations at $C=5-7$ were found to be well converged, but as turbulence remained stronger for $C=8$, convergence was poor.
Following the success of the short-time optimisation in the shear turbulence regime, we performed short-$\mathcal{T}$\,optimisations at $C=8$. 
In this case, an intermediate value of $\mathcal{T}=200$ was sufficient to achieve good convergence. Based on the observations for optimisations in the shear turbulence regime, we consider optimisation in the reduced space $O_{Fm}$.  Figure \ref{Nu-A0-C08}(\textit{a}) shows $\Nu(t)$ for $O_{F4}$ as an example. 
The target time $\mathcal{T}$ is not sufficient to capture the statistics of the endstate, but is sufficient to capture the initial response to the force.  With an increase in force amplitude, the Nusselt number gradually increases.  
Figure \ref{Nu-A0-C08}(\textit{b}) compares $O_{Fm}$ optimised for the unsteady convective turbulence versus optimisation for the artificially stabilised steady two-dimensional laminar state (setting $K=1$ as before). 
Unlike for shear-turbulence,
this time they show little difference, and the  optimal forces are close in structure, similar to figure \ref{OF-m-C01}(\textit{i-m}) -- when the flow is not too chaotic, including time dependence in the optimisation does not improve the Nusselt number further. 

\begin{figure}
         	\centering
         	\includegraphics[angle=0,width=0.9\textwidth]{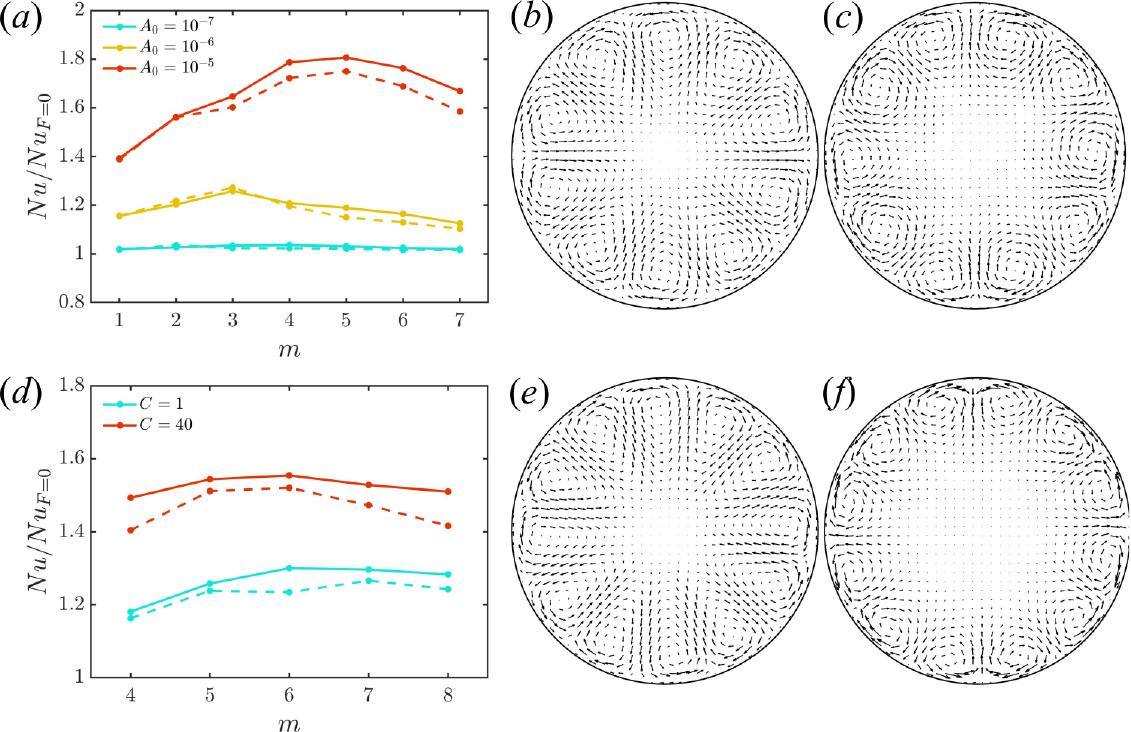}  
         	\caption{ Optimisation at larger parameters.  
            Top row: $C=16$, $\Rey=3000$, $A_0$ in key to (\textit{a}),
            $O_{F5}$ for steady and unsteady flow in (\textit{b}) and (\textit{c}).
            \AW{
            Bottom row: $\Rey=5000$, $A_0=10^{-5}$, $C$ in key to (\textit{d}),
            $O_{F6}$ at $C=1$ for steady and unsteady flow in (\textit{e}) and (\textit{f}).
            }
            In (\textit{a,d}), solid is for the  
            turbulent state, and dashed line indicates a force optimised with a steady state. 
            The largest arrow has magnitude $1.4 \times 10^{-3}$ in $(\textit {b})$, $1.9 \times 10^{-3}$ in $(\textit {c})$,$1.7 \times 10^{-3}$ in $(\textit {e})$ and $2.4 \times 10^{-3}$ in $(\textit {f})$.
            }
         	\label{Nu-A0-C16}
        \end{figure}

    Towards more chaotic convective turbulence, a further optimisation was carried out at $C=16$.  
    Due to the stronger chaos, the target time was again reduced to $\mathcal{T}=100$.  Comparison between $O_{Fm}$ optimised with the short $\mathcal{T}$ (solid line) and a steady two-dimensional laminar state (dash line) is shown in figure \ref{Nu-A0-C16}(\textit{a}). 
    Improvement for the short-$\mathcal{T}$
    optimisation starts to be seen at larger $A_0=10^{-5}$.   
  The roll structures of the force optimised
 with the short $\mathcal{T}$ 
 at all amplitudes are found to be located closer to the wall than for the corresponding steady calculation.  
  $O_{F5}$ at $A_0=10^{-5}$ for the two optimisations are
 shown in figure \ref{Nu-A0-C16}(\textit{b,c}).
 Such roll structures improve $\Nu$ at a larger force amplitude, similar to  optimisation in the shear-turbulence regime.  Similar conclusions were drawn at $C=32$.

 \CSJ{Further optimisations were also performed at $Re=5000$, $C=1$ and $40$,  corresponding to the strong shear-turbulence and convective-turbulence regimes (see figure \ref{fig:SCL}).  A short target time $\mathcal{T}=40$ was necessary for optimisations with unsteady flows in both cases, due to very strong chaos. 
 Comparison between 
 $\Nu$ for $O_{Fm}$ at $A_0=10^{-5}$ optimised with the short $\mathcal{T}$ (solid line) and when imposing a steady two-dimensional laminar state during the optimisation (dashed line) are shown in figure \ref{Nu-A0-C16}(\textit{d}). The inclusion of time-dependence leads to an improvement in $\Nu$ in both cases at $C=1$ and $C=40$. 
 The roll structures of the force optimised using unsteady flow are found to be closer to the wall than when steady flow is imposed via suppression of streamwise dependence (figure \ref{Nu-A0-C16}(\textit{e,f})).
 }

\AW{In this section we have found that the added force can laminarise convective turbulence, where it returns to a streamwise independent state.  Pushing $\Rey$ and $C$ to larger values, such that the flow is chaotic, further enhancement in $\Nu$ is found when time-dependence is included in the optimisation, and rolls are again found to be located closer to the wall. }
 
\section{Conclusions}
  \label{sec:Conclusions}

  In this work we have developed a heat transfer optimisation method, based on a variational technique \citep{pringle2010using, marensi2020designing}, designed to identify the optimal body force that maximises heat transfer, 
  in particular, 
  \AW{subject to the time-dependent governing equations for the flow and subject to limited amplitude of the force}.
 Focussing primarily on the feasibility and practicality of the method, optimisations have been conducted mostly at $Re=3000$ with the constant temperature boundary condition,
 but the method has been applied to flows initially in each of the typical states of vertical heated pipe flow, i.e.\ 
  shear turbulence ($C=1$) 
  laminarised flow ($C=3$), 
  and convective turbulence ($C\ge 4)$. 
  
Our optimisations
confirm that the optimal body force is 
predominantly
characterised
by near-wall vortex structures that are uniform
  in the streamwise direction.  
  This is consistent with optimised steady velocity fields that have been computed, e.g.\ \cite{meng2005field}.
\AW{Direct comparison with previous results is not straight-forward, however, since the optimisations were for the velocity field, rather than for a force, 
and the velocity field \EM{was}
independent of a buoyancy term (i.e.\ independent of $C$).
Although such optimisations were not subject to a momentum 
equation for the velocity, the velocity was found to satisfy an equation similar to the Navier-Stokes equation with the addition of the `synergy force', of the form $\vec{F}_s=c\,A\,\bnabla T+\vec{u}\cdot\bnabla\vec{u}$,
in which $\vec{u}$ is time-independent.
The scalar field $A(\vec{x})$ is a Lagrange multiplier for the heat equation, and the constant $c$ is a free parameter that is inversely proportional to a Lagrange multiplier of the viscous dissipation of the velocity field.  
The forces added to the Navier-Stokes in the present case are $(4/\Rey)C\,T\hat{\vec{z}}+\vec{F}$.  
Whilst we have made comparisons with optimisations we have obtained for $\vec{F}$ with a steady response velocity,
it should be cautioned that this is not directly equivalent to comparing with $\vec{F}_s$, the latter of which may not exist if subject to limited magnitude, as only $c$ can be varied.  
Viscous dissipation is not considered in our calculations, while unrestricted dissipation corresponds to $c\to\infty$ in the synergy approach.
The novelty and advantage of our approach is that $\vec{u}$ is subject to the (steady/unsteady) governing equations, where $\vec{u}$ may be significantly affected by the flow regime determined by $C$, an optimal $\vec{F}$ exists for arbitrarily small amplitude $A_0$, which changes with variation in $A_0$, and the target time $\mathcal{T}$ is also an extra parameter in our calculations.}

  Forces optimised with different target times were found to exhibit different rotational symmetry. Specifically, the short-time optimal forces corresponded to larger azimuthal wave numbers, while 
  \AW{for $\mathcal{T}\gtrsim 400$ the optimal force is unique} with smaller azimuthal wavenumber.
   This flow pattern aligns with linear optimal perturbations that aim to maximise flow perturbation growth in isothermal flow \citep{schmid2007nonmodal}.
  Forcing such modes efficiently modifies the flow.
\AW{An implication of this result is that if there is limited time to impel a change in the velocity field, or if we link reduced response time with passage through a short section of pipe, then larger $m$ might be favoured.}

In the laminarisation regime  ($C=3$), 
heat transfer increases with force amplitude $A_0$, as expected, then increases significantly at the point at which time-dependent flow is triggered by the force.
To reduce computational costs considerably,
we opted to compare the Nusselt number of flows forced by optimals constrained in the Fourier space, $O_m$ and $O_{Fm}$ (the former keeping modes $0,m,2m,...$ and the latter only keeping modes $0$ and $m$).
This method is found to be efficient
and permits examination upon dependence on  $m$.  (Optimisation in the full space $O_1$ leads to an $O_m$ with $m$ unspecified.)
For very small $A_0$, rotational symmetry $m=1$ is favoured, then $m$ increases with $A_0$.  
However, rather than the enhancement in $\Nu$ spiking for a particular $m$, the enhancement is similar over a broad range in $m$.
Also for the laminarisation regime,
there is a jump up in $\Nu$ when turbulence is first triggered, but for larger $A_0$ turbulence does not necessarily lead to an increase in $\Nu$. 
Visualisations of heat flux reveal that disorganisation caused by unsteadiness of the rolls inhibits heat transfer.  
\AW{An implication if this result is that 
an approach that seeks to simply trigger turbulence may not be efficient in enhancing $\Nu$.  Organised rolls may be better than turbulent rolls, depending on the amplitude of roll that can be achieved.
}


Optimisation in the shear-turbulence regime ($C=1$) is challenging, as the flow is highly time-dependent and chaotic, preventing the long target times \AW{$\mathcal{T}\gtrsim 400$ used in the laminar case}. 
 However, the method is found to still be effective for much shorter times, with $\mathcal T = 50$.
\AW{Optimisations for unsteady and steady flow have been compared. (Short $\mathcal{T}$ for the former, for the latter, imposing axial-independence results in a
flow that is steady in response to the force that is captured with a long $\mathcal{T}$.)}
It is found that including time-dependence results in a force with rolls located closer to the wall and leads to flows with greater $\Nu$.
Similarly, in the more chaotic convective state at $C=16\,,32; \Rey=3000$, \CSJ{and the flow at $C=1 \,, 40 \,; \Rey=5000$,} optimisations with short $\mathcal{T}$ show roll structures closer to the wall.
\AW{Therefore, the location of the vortex is another consideration for potential application.}

While it is acknowledged that accurately inducing the desired flow 
in practice is challenging, optimisations under the laminar steady state assumption, such as those by \cite{meng2005field}, have inspired designs like the alternating elliptical axis tube, discrete double-inclined ribs tubes \citep{li2009turbulent} and many other applications \citep{liu2013comprehensive,sheikholeslami2015review}. 
\AW{
Our results suggest that 
additional factors are worthy of consideration -- the 
distance of the rolls from the wall
affects the heat transfer, turbulence can both enhance or disrupt heat transfer.
These will also be affected by the flow regime, 
here influenced by the alignment of buoyancy with the vertical pipe axis.
Response time is another parameter, which may affect the choice of $m$ in this case.
}



An important consideration that we have not directly included in our optimisation is the associated pumping power required for the flow, nor the power expended by the force.  
\AW{For our calculations, the proportional increase in wall friction and $\Nu$ are similar, but it should be noted that simply increasing the pressure gradient does not lead to a proportional increase in heat transfer.  Indeed, for laminar flow, $\Nu$ is independent of $\Rey$ \citep{su2000linear}.  Table \ref{tab:PEC} shows the energy consumption and a performance evaluation criterion (PEC) close to the popular measure
developed by \cite{webb1972application},
 used in \cite{meng2005field} and the more recent review of \cite{li2023heat}.
 We have replaced the friction ratio measured by $\dot{W}_P/\dot{W}_{P,F=0}$ with
 $(\dot{W}_P+\dot{W}_F)/\dot{W}_{P,F=0}$ in order to include the work done by the added force.
The calculation using steady flow overestimates the enhancement, by 16\% and 31\% for the relatively quiescent cases at $\Rey=3000,\,C=3$ and $C=8$, and more substantially when the flow is chaotic, by 55\% and 57\% at $C=1$ (shear turbulence) and $C=16$ (strong convection-driven turbulence). It is overestimated by 126\% for the case at $\Rey=5000,\,C=1$.
}

\AW{
In summary, including the \EM{full} governing equations for the velocity provides a greater range of parameters that are found to influence the optimal force configuration.
Further including time-dependence in the optimisation is challenging, but 
can lead to further enhancement of the heat transfer and significantly improves reliability in prediction of the overall heat transfer enhancement. 
}


\begin{table}
\begin{center}
\def~{\hphantom{0}}
\begin{tabular}{cl|llll|l}
& &
\multicolumn{3}{c}{$Re=3000$}  &              & $Re=5000$ \\
\hline
\multicolumn{1}{l}{}      & $C$               & 1        & 3       & 8        & 16  & 1 \\ 
\hline
 Unforced & $\dot{W}_{P,F=0}$ & $3.63\times10^{-2}$ &$2.09\times10^{-2}$ &$1.78\times10^{-2}$ & $1.13\times10^{-2}$ &$3.46\times 10^{-2}$  \\ 
\hline
\multirow{4}{*}{Steady}   & $\dot{W}_{P}$        & $5.24\times10^{-2}$  & $4.65\times10^{-2}$ & $3.54\times10^{-2}$  & $1.93\times10^{-2}$ &$4.23\times 10^{-2}$ \\ 
                          & $\dot{W}_{F}$        & $6.93\times10^{-4}$ & $6.65\times10^{-4}$ & $6.61\times10^{-4}$  & $4.3\times10^{-3}$  &$7.08\times 10^{-4}$ \\ 
                          & $Nu/Nu_{F=0}$ (pred.) &1.76 &3.04 &2.69 &2.18 &1.52
                          \\ 
                          & $Nu/Nu_{F=0}$ (obs.)& 1.49     & 2.75    & 2.29     & 1.75     &1.23  \\ 
                          & PEC             & 1.31     & 2.10    & 1.81     & 1.37     &1.14 
                          \\
                          \hline 
\multirow{4}{*}{Unsteady} & $\dot{W}_{P}$        & $5.52\times10^{-2}$  & n/a      & $3.61\times10^{-2}$ & $2.04\times10^{-2}$  &$4.37\times10^{-2}$ \\ 
                          & $\dot{W}_{F}$        & $2.40\times10^{-3}$  & n/a      & $1.2\times10^{-3}$  & $1.5\times10^{-3}$ &$1.7\times10^{-3}$\\ 
                          & $Nu/Nu_{F=0}$& 1.57     & n/a      & 2.34     & 1.81     &1.30 \\ 
                          & PEC             & 1.34     & n/a      & 1.83     & 1.45  &1.19  
\end{tabular}
 \caption{\CSJ{Energy consumption of flow forced by $O_{F5}$ at $\Rey=3000$ and $O_{F6}$ at $\Rey=5000$, with
 $A_0=10^{-5}$ and several $C$.
 $\dot{W}_P=\langle \frac{4}{Re}(1+{\beta^{'}})u_{tot,z} \rangle$ is the pumping power. $\dot{W}_F=\langle|\vec{F}\cdot \vec{u}_{tot}| \rangle$ is rate of the work done by the optimal force. For the `Steady' case, the flow is restricted to be axially-independent during the optimisaton, which leads to steady flow and a `predicted' $\Nu$.  The force is then applied in a DNS to calculate the `observed' $\Nu$.
 For the `Unsteady' case, the flow is three-dimentionsal and time-dependent in the optimisation. 
 The performance evaluation criteria (PEC) is defined  $\mathrm{PEC}=({Nu/Nu_{F=0}})/{((\dot{W}_P+\dot{W}_F)/\dot{W}_{P,F=0})^{\frac{1}{3}}}$.
 Flow is laminarised (steady) for the case marked `n/a'.
 }
     }
     \label{tab:PEC}
     \end{center}
\end{table}


\backsection[Acknowledgements]{ This work used the Cirrus UK National Tier-2 HPC Service at EPCC (http://www.cirrus.ac.uk) funded by the University of Edinburgh and EPSRC (EP/P020267/1).} 

\backsection[Funding]{ S.C. acknowledges the funding from Sheffield–China Scholarships Council PhD Scholarship Programme (CSC no. 202106260029).}

\backsection[Declaration of interests]{
The authors report no conflict of interest.}


\backsection[Author ORCIDs]
{Shijun Chu, https://orcid.org/0000-0002-3037-6370;
 Ashley P. Willis, https://orcid.org/0000-0002-2693-2952;
 Elena Marensi, https://orcid.org/0000-0001-7173-4923.}


\appendix

\section{\CSJ{Scalar dissipation} }\label{appA}
We start from the dimensionless governing equation for temperature
 \begin{eqnarray}
        	\frac{\partial T }{\partial t}+({\vec{u}_{tot}}\bcdot \vec{\nabla}) T &=&\frac{1}{Re\,Pr}\nabla ^{2}T -{\vec{u}_{tot}}\bcdot{\hat{\vec{z}}}\,a_{tot}(t) \, .
        	\label{T-total-appen}
\end{eqnarray} 
Multiplying (\ref{T-total-appen}) by the temperature $T$
and integrating in time from $0$ to $\mathcal{T}$,
we obtain a new balance equation for $T^2$,
\begin{eqnarray}
        	 \int_{0}^{\mathcal{T}} \frac{1}{2} \left(
             \frac{\partial}{\partial t} T^2 + 
             ({\vec{u}_{tot}}\bcdot \vec{\nabla}) T^2
             \right)\mathrm{d}t  & = &
             \int_{0}^{\mathcal{T}} \left(
             \frac{T}{Re\,Pr}\nabla ^{2}T - T\,{\vec{u}_{tot}}\bcdot{\hat{\vec{z}}}\,a_{tot}(t)\right)\mathrm{d}t  \, .
        	\label{T-total-apenT}
\end{eqnarray} 
We aim for a statistically steady state, and therefore drop the first term on the left hand side.
Integrating (\ref{T-total-apenT}) over the entire domain, the second term also vanishes, as it transports $T^2$ but conserves the total. Applying integration by parts on the first term on the right hand side term leads to
 \begin{eqnarray}
      0 &\approx&
        	\int_{0}^{\mathcal{T}} \left( \frac{1}{Re Pr}\langle  \vec {\nabla} \cdot (T\vec {\nabla} T) - (\vec {\nabla} T)^2 \rangle - \langle T\,{\vec{u}_{tot}}\cdot{\hat{\vec{z}}}\,a_{tot}(t) \rangle \right)\mathrm{d}t \, .
        	\label{T-volume}
\end{eqnarray}
Note the `dissipation' term involving $\langle(\bnabla T)^2\rangle$ can only lead to a reduction in $T^2$.
Pulling this term across and 
applying the divergence theorem gives
\begin{eqnarray}
            \int_{0}^{\mathcal{T}}  \frac{1}{Re\, Pr}\langle (\vec{\nabla} T)^2\rangle \, \mathrm{d}t
            \approx
            \int_{0}^{\mathcal{T}} \left( \oint  \frac{1}{Re\, Pr}T\,\vec{\nabla}T\cdot\vec{\mathrm{d}S} 
            -
            \langle T\,{\vec{u}_{tot}}\cdot{\hat{\vec{z}}}\,a_{tot}(t) \rangle 
            \right)
            \, \mathrm{d}t 
 \,.
        	\label{T-volume1}
\end{eqnarray}
The surface integral term can be rewritten, given that $T$ is constant at the wall (and
applying periodicity at the ends) 
\begin{eqnarray}
        \oint\frac{1}{Re\, Pr}T\,\vec{\nabla}T\cdot\vec{\mathrm{d}S} ~=~ T|_{r=1}\,\int\frac{1}{Re\, Pr}\vec{\nabla}T\cdot\vec{\mathrm{d}S}&=&T|_{r=1}\,Q_{in}\, ,
        	\label{T-Qin}
\end{eqnarray}
where $Q_{in}$ is the total heat flux, entering at the wall.
For the heat sink term in (\ref{T-volume1})  we make the further approximation that  
$\langle T\,{\vec{u}_{tot}}\bcdot{\hat{\vec{z}}}\,a_{tot}(t)\rangle \approx T_b \, \langle {\vec{u}_{tot}}\bcdot{\hat{\vec{z}}}\,a_{tot}(t)\rangle = 
T_b \,Q_{out}$,
where $Q_{out}$ represents the total net rate at which heat is carried out of the domain.
(Observe that for a spatially uniform heat sink $\epsilon$, as in \cite{marensi2021suppression}, no approximation is necessary: 
$\langle T\,\epsilon(t)\rangle = T_b \, \langle \epsilon(t) \rangle = 
T_b \,Q_{out}$.  While that models a temporal rather than axial change in temperature, very little difference is observed, see figure \ref{fig:SCL}({\it b}).  Note that the velocity profile tends to be `flatter', i.e.\ providing a more uniform heat sink, when the flow is turbulent.)
Assuming that on average $Q_{in}=Q_{out}$,
(\ref{T-volume1}) becomes
 \begin{eqnarray}
        	\int_{0}^{\mathcal{T}}  \frac{1}{Re\, Pr}\langle(\vec{\nabla} T)^2\rangle \,  \mathrm{d}t\approx\int_{0}^{\mathcal{T}}   (T|_{r=1}-T_b) \, Q_{in}  \, \mathrm{d}t\,.
        	\label{Entransy-final}
\end{eqnarray}
This equation implies the maximisation (minimisation) principle for $\langle(\bnabla T)^2\rangle$ -- 
when the temperature difference is fixed, maximising the left hand side implies the largest $Q_{in}$, and hence the largest $\Nu$.
(When $Q_{in}$ is fixed, minimising the left hand side implies the smallest temperature difference, and hence the largest $\Nu$.)

\bibliographystyle{jfm}
\bibliography{jfm2esam}


\end{document}